\documentclass[usenatbib]{mnras}
\usepackage{graphicx, rotating, longtable, lscape, amsmath}

\newcommand\msun{{\rm M_{\odot}}}

\def\go{
\mathrel{\raise.3ex\hbox{$>$}\mkern-14mu\lower0.6ex\hbox{$\sim$}}
}
\def\lo{
\mathrel{\raise.3ex\hbox{$<$}\mkern-14mu\lower0.6ex\hbox{$\sim$}}
}
\title[{\em Swift}-XRT observations during O3]
{{\em Swift}-XRT follow-up of gravitational wave triggers during the third aLIGO/Virgo observing run}
\author[K.L. Page et al.]{K.L. Page$^{1}$, P.A. Evans$^{1}$, A. Tohuvavohu$^{2}$, J.A. Kennea$^{3}$, N.J. Klingler$^{3}$,\newauthor S.B. Cenko$^{4,5}$, S.R. Oates$^{6}$, E. Ambrosi$^{7}$, S.D. Barthelmy$^{4}$, A.P. Beardmore$^{1}$,  \newauthor M.G. Bernardini$^{8}$, A.A. Breeveld$^{9}$, P.J. Brown$^{10}$, D.N. Burrows$^{3}$, S. Campana$^{8}$, \newauthor R. Caputo$^{4}$, G. Cusumano$^{7}$, A. D'A{\` i}$^{7}$, P. D'Avanzo$^{8}$, V. D'Elia$^{11,12}$,  \newauthor M. De Pasquale$^{13}$, S.W.K. Emery$^{9}$, P. Giommi$^{12}$, C. Gronwall$^{3,14}$,   D.H. Hartmann$^{15}$, \newauthor H.A. Krimm$^{16}$,  N.P.M. Kuin$^{9}$, D.B. Malesani$^{17}$, F.E. Marshall$^{4}$, A. Melandri$^{8}$,  \newauthor J.A. Nousek$^{3}$, P.T. O'Brien$^{1}$, J.P. Osborne$^{1}$, C. Pagani$^{1}$, M.J. Page$^{9}$,  \newauthor D.M. Palmer$^{18}$, M. Perri$^{12,11}$, J.L. Racusin$^{4}$, T. Sakamoto$^{19}$, B. Sbarufatti$^{3}$,  \newauthor J.E. Schlieder$^{4}$, M.H. Siegel$^{3}$, G. Tagliaferri$^{8}$ \& E. Troja$^{4,20}$ \\
$^{1}$ School of Physics and Astronomy, University of Leicester, LE1 7RH, UK\\
  $^{2}$  Department of Astronomy and Astrophysics, University of Toronto, Toronto, ON, Canada\\
  $^{3}$  Department of Astronomy and Astrophysics, The Pennsylvania State University, University Park, PA 16802, USA\\
  $^{4}$ Astrophysics Science Division, NASA Goddard Space Flight Center, Greenbelt, MD 20771, USA\\
  $^{5}$ Joint Space-Science Institute, Computer and Space Sciences Building, University of Maryland, College Park, MD 20742, USA\\
 $^{6}$  School of Physics and Astronomy, University of Birmingham, B15 2TT, UK\\  
  $^{7}$ INAF -- IASF Palermo, Via Ugo La Malfa 153, I-90146, Palermo, Italy\\
  $^{8}$ INAF -- Osservatorio Astronomico di Brera, Via Bianchi 46, I-23807 Merate, Italy\\
  $^{9}$ University College London, Mullard Space Science Laboratory, Holmbury St. Mary, Dorking, RH5 6NT, UK\\
  $^{10}$ George P. and Cynthia Woods Mitchell Institute for Fundamental Physics and Astronomy, Mitchell Physics Building, \\Texas A.\&M. University, Department of Physics and Astronomy, College Station, TX 77843, USA\\
  $^{11}$ INAF-Osservatorio Astronomico di Roma, Via Frascati 33, I-00040 Monteporzio Catone, Italy\\
  $^{12}$ Space Science Data Center (SSDC) - Agenzia Spaziale Italiana (ASI), I-00133 Roma, Italy\\
  $^{13}$  Department of Astronomy and Space Sciences, Istanbul University, Beyaz{\i}t 34119, Istanbul, Turkey\\
  $^{14}$  Institute for Gravitation and the Cosmos, The Pennsylvania State University, University Park, PA 16802, USA\\
  $^{15}$  Department of Physics and Astronomy, Clemson University, Kinard Lab of Physics, Clemson, SC 29634-0978, USA\\
  $^{16}$  National Science Foundation, Alexandria, VA 22314, USA\\
  $^{17}$  DTU Space, National Space Institute, Technical University of Denmark, Elektrovej 327, 2800 Kongens Lyngby, Denmark\\
  $^{18}$ Los Alamos National Laboratory, B244, Los Alamos, NM, 87545, USA\\
$^{19}$  Department of Physics and Mathematics, Aoyama Gakuin University, Sagamihara, Kanagawa, 252-5258, Japan\\
  $^{20}$  Department of Physics and Astronomy, University of Maryland, College Park, MD 20742, USA\\
}

\date{Accepted XXX. Received YYY; in original form ZZZ}

\pubyear{2020}

\begin{document}
\label{firstpage}
\pagerange{\pageref{firstpage}--\pageref{lastpage}}

\maketitle

\begin{abstract}

  The {\em Neil Gehrels Swift Observatory} followed up 18 gravitational wave (GW) triggers from the LIGO/Virgo collaboration during the O3 observing run in 2019/2020, performing approximately 6500 pointings in total. Of these events, four were finally classified (if real) as binary black hole (BH) triggers, six as binary neutron star (NS) events, two each of NSBH and Mass Gap triggers, one an unmodelled (Burst) trigger, and the remaining three were subsequently retracted. Thus far, four of these O3 triggers have been formally confirmed as real gravitational wave events. While no likely electromagnetic counterparts to any of these GW events have been identified in the X-ray data (to an average upper limit of 3.60~$\times$~10$^{-12}$ erg~cm$^{-2}$~s$^{-1}$ over 0.3--10~keV), or at other wavelengths, we present a summary of all the {\em Swift}-XRT observations performed during O3, together with typical upper limits for each trigger observed. The majority of X-ray sources detected during O3 were previously uncatalogued; while some of these will be new (transient) sources, others are simply too faint to have been detected by earlier survey missions such as {\em ROSAT}. The all-sky survey currently being performed by {\em eROSITA} will be a very useful comparison for future observing runs, reducing the number of apparent candidate X-ray counterparts by up to 95~per~cent.

\end{abstract}


\section{Introduction}
\label{intro}

Gravitational waves (GWs) were an important prediction of Einstein's 1915 General Theory of Relativity. 
Experiments to try and detect them were first pioneered back in the 1960s, with the first real steps towards LIGO (Laser Interferometer Gravitational-Wave Observatory) taken in the 1980s\footnote{See https://www.ligo.caltech.edu/page/timeline for a brief history of LIGO.}. LIGO was inaugurated in the final quarter of 1999, with science runs starting in 2002 \citep{abbott04}. Construction of the European Virgo project started in 1996, with the initial detector being completed in 2003. Science runs began in 2007, with a joint data analysis agreement with LIGO\footnotemark[1]. The first follow-up of LIGO alerts by the {\em Neil Gehrels Swift Observatory} \citep[{\em Swift} hereafter;][]{geh04} occurred in 2010, though these two events were not astrophysical in origin \citep{evans12}.

`Advanced LIGO' \citep[aLIGO;][]{ligo15} began its first observing run (termed O1) on 2015 September 12, running until 2016 January 19, and yielding three triggers in that time. Of these, two (GW~150914 and GW~151226) were identified as binary black hole (BBH) mergers \citep{abbott16a, abbott16b}, with the third (G194575) being classified as a noise event. In addition, later offline analysis also suggested that trigger LVT~151012, while a lower significance detection, was still 87~per~cent likely to be of astrophysical origin \citep{abbott16c}. 

Following an upgrade and commissioning period, the second aLIGO observing run, O2, began on 2016 November 30, running until 2017 August 25. At the very end of this interval, from 2017 August 01, the Advanced Virgo detector \citep{virgo15} also joined the run, allowing for three detector observations of GW events. During O2, a further eight confident GW triggers were identified \citep{abbott19} -- seven BBHs, and one binary neutron star (BNS) merger, the latter leading to the first detection of an electromagnetic (EM) counterpart of a GW event \citep[e.g., ][]{abbott17, coulter17, evans17, goldstein17, hallinan17, pian17, troja17}.

The first part of the third observing run, O3a, ran from 2019 April 01 until 2019 September 30, at which time a one month commissioning break took place. O3b commenced on 2019 November 01, with the initial plan to run until 2020 April 30. However, due to the COVID-19 pandemic, the observing run was ended early, on March 27. O3 was the first observing run where triggers were publicly announced in real time, with details available online at https://gracedb.ligo.org/superevents/public/O3/. KAGRA \citep[Kamioka Gravitational Wave Detector; ][]{kagra}, Japan's GW observatory, began real-time observations in 2020 March, although the interferometer was not sensitive enough during the final weeks of O3 to detect the same GW events as LIGO and Virgo.

The discovery of GWs has opened up a new window on the cosmos, allowing astronomers to investigate sources which emit little, or no, light (and would therefore be invisible to traditional telescopes), and to delve into some of the most extreme environments conceivable -- the merging of BH and/or NS.
While the initial detection of these events requires large GW interferometers, it is the subsequent follow-up by other observatories which can lead to good localisations, and details about temporal and spectral evolution of any afterglow. With its rapid response capability (typical slew times of $\lo$~100~s), together with co-aligned X-ray and UV/optical telescopes, {\em Swift} is well placed to add to these observations. While there are many transients in the Universe, most of which will be unrelated to GW events (see discussion in Section~\ref{agn}), there are far fewer serendipitously detected in X-rays than at optical wavelengths. Searches for optical counterparts to O3 triggers have been published by \cite{gompertz20}, \cite{antier20} and \cite{kasliwal20b}, among others.
In addition, a decaying X-ray afterglow is a distinguishing feature of many Gamma-Ray Bursts afterglows \citep[GRBs; e.g.,][]{nousek06}, with short GRBs expected to be formed during NS merger events, alongside the gravitational waves \citep{eichler89, mochkovitch93}. The X-ray bandpass is therefore a useful and interesting region to search.

\cite{evans16c} summarised the {\em Swift} follow-up of LIGO triggers in O1, while \cite{klingler19} presented the same for O2. In this paper, we cover the {\em Swift} X-ray observations from O3. A companion paper by Oates et al. (in prep.) will present the corresponding optical and ultraviolet (UV) data from {\em Swift}, and a paper combining the $\gamma$-ray data from {\em Swift} and {\em Fermi} is also planned. We refer the reader to \cite{evans16c} and \cite{klingler19} for details of the X-ray data processing, analysis and source detection \citep[see also][]{1sxps, 2sxps}, which did not significantly change for O3. Throughout the paper, upper limits are given at the 3$\sigma$ level. Magnitudes are provided in the AB system.

\section{Swift observations}
\label{swift}

The {\em Swift} observatory, launched in 2004, comprises three instruments: the wide-field ($\sim$~2~sr) Burst Alert Telescope \citep[BAT; ][]{scott05}, sensitive to 15--350~keV; the X-ray Telescope \setcitestyle{square}\citep[XRT, 0.3--10~keV, with a circular field of view (FOV) of diameter 23.6~arcmin;][]{bur05}; and the UV/Optical Telescope \setcitestyle{round}\citep[UVOT, with seven filters spanning $\sim$1700--6000 \AA, and a square FOV 17~arcmin each side;][]{rom05}.

{\em Swift} was designed to detect and rapidly follow up GRBs. While excellent at this job, its remit has expanded over the years, with the satellite becoming the go-to mission for any X-ray or UV transient source, especially where rapid observations are required; since 2015, this has included the search for EM counterparts to GW events. The best case scenario would obviously be for {\em Swift}-BAT to trigger independently on a short GRB corresponding to a GW trigger, allowing the observatory to localise the source to a few arcmin (from BAT) or (sub)arcsec (UVOT/XRT) promptly and autonomously. Unfortunately, such a situation will be rare, and has not yet occurred; while GRB~170817A, the short burst associated with GW~170817, triggered the {\em Fermi} Gamma-ray Burst Monitor \citep[GBM; ][]{21506, goldstein17} and was detected by INTEGRAL \citep{integral}, the location of the event was occulted by the Earth for {\em Swift} \citep{evans17}. 
Therefore, since the error regions of the LIGO GW detections are typically hundreds of square degrees in area, very much larger than the fields of view of the XRT or UVOT, a method to optimise {\em Swift}'s ability to tile large areas of the sky was put in place \citep{evans16a}.

As mentioned by \cite{evans16b,evans16c}, the planned large-scale rapid tiling ability had not been commissioned by the start of O1. On 2016 January 13 an initial test of this rapid tiling was scheduled to observe GW~151226 \citep{evans16c}, showing that the spacecraft could safely perform hundreds of short ($\sim$~60~s) exposures in quick succession. This new observing mode, fully operational by the time of of O2, allows {\em Swift} to cover $\sim$50~deg$^2$ per day, a substantial increase over the possible response in O1. By the start of O3, work had been undertaken to optimise the scheduling of tiles for both efficiency and spacecraft safety \citep[see discussion in][]{aaron18, aaronjamie18}. In addition, further progress had been made such that these tiling observing plans could be uploaded more easily, allowing detailed follow-up by {\em Swift} to begin even more quickly.

Between each of the LIGO observing runs, the plan for {\em Swift} follow-up was optimised, based on lessons learned from the previous data. LIGO performs searches for two different types of event: Compact Binary Coalescence (CBC) and unmodelled (Burst) triggers\footnote{https://emfollow.docs.ligo.org/userguide/analysis/searches.html}. The CBC triggers are modelled searches, looking specifically for BNS, NSBH and BBH mergers, while the unmodelled triggers have no prior assumptions regarding the signal, and could be caused by different astrophysical events (for example core-collapse supernovae).

For {\em Swift} observations of all CBC triggers, the LVC\footnote{Throughout this paper, LVC is used as an abbreviation for publications by the LIGO Scientific Collaboration and the Virgo Collaboration.} (LIGO Scientific Collaboration and Virgo Collaboration) probability map was convolved with an appropriate 3D galaxy catalogue (that is, the predicted distance information of the merger from Earth was included in the calculation), to account for the fact that such CBC events (the majority of LIGO triggers) are expected to occur in nearby galaxies. Further details of this process are given in \cite{evans16b} and \cite{klingler19}. For all O3 CBC triggers, the Two Micron All Sky Survey (2MASS) Photometric Redshift Catalog \citep[2MPZ;][]{bilicki14} was used. For unmodelled triggers, a frequency of $>$1~kHz could correspond to events in our own Galaxy; in those specific cases, convolution with the Galactic plane was performed instead. For the lower frequency unmodelled events, the Gravitational Wave Galaxy Catalog \citep[GWGC;][]{white11} was implemented, since these events are expected only to be detectable out to $\sim$~100~Mpc \citep{abbott19b}, and the GWGC data are more complete than 2MPZ in this regime.

In previous observing runs, the tiling pattern performed by {\em Swift} has been based entirely around the XRT FOV, since it is a little larger than the UVOT field \cite[see fig. 3 of][]{evans16c}. While this avoids time wasted in overlapping sky area, it also leads to there being areas observed by XRT, but outside the UVOT FOV. Given that the EM counterpart to GW~170817 was detected by the UVOT, not XRT \citep{evans17}, it was concluded that the tile selection criteria should be modified such that fields containing potential host galaxies can be offset or split, ensuring that galaxies fall entirely within the UVOT FOV \citep[see also][]{klingler19}.

\cite{klingler19} details the follow-up criteria applied to triggers in O2 for the {\em Swift} observations. One change implemented at the start of O3a was that, for a BBH or Mass Gap\footnote{A Mass Gap trigger implies a system with at least one compact object whose mass is in the hypothetical `mass gap' between NS and BH, taken to be 3--5~$\msun$.} trigger, we required that the minimum area enclosing 90 per~cent of the probability in the convolved sky map be $\lo$ 10~deg$^2$. Previously the constraint had depended on fraction of the LVC probability region contained within the 400 most probable XRT fields. That is, given that the likelihood of the merger of two black holes leading to an EM counterpart is low \citep{kam13, metzger19}, it was decided that {\em Swift} would only follow up well-localised BBH events.  One exception was made for S190414m, when a new version of the tiling software was tested; for this source, the area enclosing 90 per~cent of the probability was 151.7~deg$^2$.

Initially, for O3a and earlier runs, any trigger which was marked as containing at least one NS (BNS or NSBH) was automatically flagged to be followed up, given that these are rarer than BBH mergers, and, from theory, more likely to lead to EM emission, in the form of a short GRB  \citep[if viewed on-axis; e.g., ][]{berger14}, and/or a kilonova, irrespective of jet alignment \citep[since such emission is more isotropic; ][]{eichler89, li98, metzger10}. However, during the month-long break between O3a and O3b, it was decided that the likelihood of the NS being {\em disrupted} should be taken into account; if a NS were simply to be swallowed whole by the companion BH, then no EM radiation would be expected. This was estimated using the equation of probabilities:

\begin{displaymath}
\begin{split}  
  P_{\rm disrupt. NS} = P_{\rm NS} \times (1-P_{\rm Terres.}) - P_{\rm NSBH}\\  + P_{\rm remnant} \times P_{\rm NSBH}
\end{split}  
\end{displaymath}

\noindent and changes were then implemented in the {\em Swift} selection algorithm on 2019 December 12. Here, P$_{\rm disrupt. NS}$ is the probability of the event containing a {\em disrupted} NS; P$_{\rm NS}$ is the probability that at least one of the compact objects is a NS (if the source is real), while P$_{\rm NSBH}$ is that of the trigger being a NSBH binary. P$_{\rm remnant}$ gives the probability that the system ejected a non-zero amount of NS material. P$_{\rm Terres.}$ signifies the probability of a trigger being of terrestrial origin -- i.e. noise. For each trigger,  P$_{\rm NS}$, P$_{\rm Terres.}$, P$_{\rm NSBH}$ and P$_{\rm remnant}$ are taken from the relevant GCN (Gamma-ray Coordinates Network) notice sent by the LVC\footnote{https://emfollow.docs.ligo.org/userguide/content.html}.
Throughout O3b there was also an additional requirement that the False Alarm Rate (FAR) needed to be $\lo$~3.17~$\times$~10$^{-9}$~Hz (i.e. less frequent than one in ten years) if P$_{\rm disrupt. NS}$~$<$~0.7; where the probability of a disrupted NS was higher, the FAR estimate was ignored. The area constraint was also rephrased in terms of the probability observable in 24 hours, P$_{\rm 24 hr}$, rather than the statement that the area enclosing 90~per~cent of the probability be $\lo$~10~deg$^2$.
Table~\ref{table:rules} summarises what was required for follow-up by {\em Swift} for each type of trigger during the O3 observing run. We do note, however, that, while we tried to follow this decision tree systematically, {\em Swift} operational constraints also needed to be taken into consideration.

Assuming the criteria were satisfied, the standard follow-up plan for {\em Swift} was dependent on the type of event announced; the default steps are summarised in Table~\ref{table:steps}, though the scheme was not always fully executed (see description of individual triggers in Section~\ref{results}). As mentioned above, the mergers of stellar mass black holes (the BBH triggers) are not generally anticipated to lead to EM radiation, so a promptly detected short GRB is not expected \citep[c.f. GW~150914; ][]{con16, con18, greiner16}; 500~s tiling observations were still performed however. When the system was thought to include a disrupted NS, or the trigger was a low-frequency unmodelled event, the observations took a two pronged approach: first rapid short tiles, repeated if the area could be covered in $<$1.5~days, followed by deeper observations. The reasoning behind this is that prompt observations would look for the rapidly fading (on-axis) afterglow (with the repeat short observations possibly serving to catch a source turning on slightly more slowly), while later ones might detect a rising (off-axis) jet \citep[see][for full details]{evans16a}. While off-axis mergers are more likely, due to simple geometric effects, the time at which the corresponding afterglow would be detectable by the XRT depends strongly on the jet parameters and the density of the circumburst medium. Observations of 500~s from three to seven days post-trigger were concluded to be a sensible compromise, following work by \cite{evans16a}. For unmodelled triggers with a frequency $>$~1~kHz, the region was to be observed for 80 s per tile continuously for four days.  These observing strategies are somewhat different from O2 \citep{klingler19}, though were not changed between O3a and O3b. In addition to this underlying plan, follow-up observations of externally-detected sources (typically optical transients) announced via the GCN would also be performed where deemed to be of interest.

\begin{table*}

  \caption{{\em Swift} follow-up criteria during O3. P$_{0.9}$ signifies 90~per~cent of the probability in the galaxy-convolved skymap; P$_{\rm 24 hr}$ signifies the galaxy-convolved probability which {\em Swift} would be able to observe in 24 hours.}

\begin{center}
\begin{tabular}{lc}
\hline
Type of trigger & Criteria \\
\hline
O3a\\
\hline
CBC; BBH & Follow if P$_{0.9}$ $\lo$ 10 deg$^2$\\
CBC; Mass Gap & Observe if P$_{\rm NS}$ $>$ 0 or P$_{0.9}$ $\lo$ 10 deg$^2$\\
CBC; NSBH & Observe all triggers with P$_{\rm NS}$ $>$ 0\\
CBC; BNS & Observe all triggers with P$_{\rm NS}$ $>$ 0\\
Unmodelled (Burst); frequency $<$ 1 kHz & Follow if P$_{\rm 24 hr}$ $>$ 0.5 \& FAR $<$ 1/yr. \\
Unmodelled (Burst); frequency $>$ 1 kHz & Follow all \\
\hline
O3b\\
\hline
CBC; P$_{\rm disrupt. NS}$ = 0 & Follow if P$_{\rm 24 hr}$ $>$ 0.5 \& FAR $<$ 1/10 yr.  \\
CBC; 0 $\lo$ P$_{\rm disrupt. NS}$ $\lo$ 0.25  & Follow if P$_{\rm 24 hr}$ $>$ 0.5 \& FAR $<$ 1/10 yr.\\
CBC; 0.25 $<$ P$_{\rm disrupt. NS}$ $\lo$ 0.7 & Follow if  P$_{\rm 24 hr}$ $>$ 0.4\\
CBC; P$_{\rm disrupt. NS}$ $>$ 0.75  & Follow if P$_{\rm 24 hr}$ $>$ 0.1\\
CBC; NSBH $>$ 0.5 &  Follow if P$_{\rm 24 hr}$ $>$ 0.75*\\
Unmodelled (Burst); frequency $<$ 1 kHz & Follow if P$_{\rm 24 hr}$ $>$ 0.5 \& FAR $<$ 1/yr.\\
Unmodelled (Burst); frequency $>$ 1 kHz & Follow all\\
\hline
\end{tabular}

\footnotesize{*This was an additional option included from 2019 November 07.}

\label{table:rules}
\end{center}
\end{table*}

\begin{table*}

  \caption{Default tiling follow-up plan for {\em Swift} observations after a GW trigger.}

\begin{center}
\begin{tabular}{ll}
\hline
Type of trigger & Steps \\
\hline
P$_{\rm disrupt. NS}$~=~0 & 500~s per field for 4 days, or until 90~per~cent of the probability is covered, whichever sooner\\
\hline
P$_{\rm disrupt. NS}$~$>$~0 & (i) 80~s tiles up to 800 fields* or until 90~per~cent of the probability had been covered, whichever sooner\\
or & (ii) If 80~s tiling completed in $<$~1.5~d, repeat until T+3d\\
Unmodelled (Burst); frequency $<$ 1 kHz & (iii) 500~s observations per field for four days\\
\hline
Unmodelled (Burst); frequency $>$ 1 kHz & 80 s per tile for four days\\

\hline
\end{tabular}

\footnotesize{* The value of 800 fields comes from simulations \citep[the population of which was based on work by][]{leo16}, which show that, in $\sim$~80~per~cent of the cases with galaxy convolution, the correct field is reached within 800 attempts; for a higher number of fields, the increase in probability per additional field observed becomes minimal.}

\label{table:steps}
\end{center}
\end{table*}

Sources detected in the XRT observations were automatically ranked, indicating how likely each one was to be the EM counterpart of the GW trigger. \cite{evans16b} provides the detailed definitions\footnote{See also https://www.swift.ac.uk/LVC/docs.php\#classes}, with Rank 1 being a candidate afterglow\footnote{To be marked as an afterglow candidate, a source must be either uncatalogued and at least 5$\sigma$ above the 3$\sigma$ upper limit from RASS or 1SXPS/2SXPS, or a known X-ray source which is 5$\sigma$ above the catalogued flux; a power-law spectrum with $\Gamma$~=~1.7 and N$_{\rm H}$~=~3~$\times$~10$^{20}$~cm$^{-2}$ is always used for the conversion between the {\em ROSAT} and {\em Swift} bands. Additionally, the source in question must lie within 200~kpc of a known galaxy (assuming it is at the distance of that galaxy). There is no requirement that the source be seen to be fading immediately.}; Rank 2 being an interesting source/possible EM counterpart\footnote{To be classed as `interesting', a source must be either uncatalogued and at least 3$\sigma$ above the 3$\sigma$ upper limit from RASS or 1SXPS/2SXPS, or fading; alternatively, it may be a known X-ray source which is 3$\sigma$ above the catalogued flux. It does not need to be near a known galaxy.}; Rank 3 indicating an uncatalogued, though faint\footnote{By faint, we mean below the 3$\sigma$ {\em ROSAT} All-Sky Survey (RASS) limit at the source position.}, source unlikely to be an afterglow; and Rank 4 corresponding to a known X-ray source not showing any unusual activity. 
To determine if a source was a known X-ray emitter, the full HEASARC X-ray Master Catalogue\footnote{https://heasarc.nasa.gov/W3Browse/all/xray.html} was consulted, as well as the 1SXPS/2SXPS \citep{1sxps, 2sxps} catalogues. In addition, comparisons with any overlapping reference fields previously observed for the {\em Swift} Gravitational Wave Galaxy Survey \citep[SGWGS;][Tohuvavohu in prep.]{klingler19} were performed. SGWGS is a pre-imaging survey of the $\sim$~14,000 most likely host galaxies for BNS mergers within $\sim$~100 Mpc, with data collected in the X-ray and UVOT ($u$ and $uvw1$ filters) bands. When complete, $\sim$~41~per~cent of the integrated luminosity within 100~Mpc will have XRT/UVOT templates with exposure times of $\sim$~1~ks with which to compare future observations.

GCN counterpart notices were initially automatically sent out for all source detected. From 2019 April 30, notices were disabled for sources of Rank 3 or 4 which had a warning flag set (see below); from 2019 May 03, Rank 4 counterpart notices were no longer sent out whether or not there was a warning flag.

Each source was then checked by a member of the XRT team, to catch any spurious detections (caused, for example, by optical loading\footnote{https://www.swift.ac.uk/analysis/xrt/optical\_loading.php}, diffuse emission or unusually high background -- all of which would raise a `warning' flag); only these `confirmed' sources are listed on the public webpage at https://www.swift.ac.uk/LVC/. In most cases, human-vetted summary GCN circulars were then sent out when observations had been completed and checked. We take this opportunity to remind users once again that the GCN {\em notices} sent out for every XRT detection are {\em automatic} and {\em preliminary}, and are aimed at informing the community as rapidly as possible of potential counterparts. Both the warning flags in the notice, and, especially, the list of confirmed sources on the website should be checked carefully before accepting the validity of the source. 

Following the tiling and initial source detection, any Rank 1 or 2 sources of interest were reobserved with high priority to check on the flux level and any variability. The default plan\footnotemark[6] called for all Rank 3 sources also to be reobserved once the tiling was completed. While these follow-up observations were sometimes performed (particularly if the source showed any sign of fading, even if at $<$2$\sigma$), the large number of the Rank 3 sources, together with other higher priority non-GW-related {\em Swift} observations, meant that this was not always the case. Rank 4 (catalogued) sources were not reobserved.

Table~\ref{table:triggers} provides the details of the triggers from O3 which {\em Swift} followed up, including the number of X-ray sources detected, while Table~\ref{table:alltriggers} lists all the O3 events, giving reasons why {\em Swift} observations were not performed where relevant.

\begin{table*}

  \caption{O3 triggers followed-up by {\em Swift}. In column 4, (B), (L) and (cWB) note which skymap area is referred to: BAYESTAR, LALInference or coherent WaveBurst respectively. Columns 5 and 6 give the fraction of this area which was covered by XRT observations of non-retracted triggers: `raw' is the full area, while `conv.' refers to the area convolved with the relevant galaxy catalogue for the CBC triggers.}

\begin{center}
\begin{tabular}{lcccccccccc}
\hline
LVC trigger &   {\em Swift} obs.  & Obs. & GW 90\% &\multicolumn{2}{c}{Frac. of skymap covered} & \multicolumn{5}{c}{Number of confirmed XRT sources}\\
ID  & start time (hr) & performed & area (deg$^2$) & raw & conv.& Total & Rank 1 & Rank 2 & Rank 3 & Rank 4\\
\hline
O3a\\
\hline
S190412m/GW190412 & T+10.6 & 97 & 156 (B)  & 0.17 & 0.17 & 3 & & & 2 & 1\\
S190425z/GW190425 &  T+4.6& 406 & 7461 (L) & 0.010 & 0.065 & 9 & & & 2 & 7\\
S190426c &  T+2.4 & 894 & 1131 (L) & 0.18 & 0.31 & 107 & & & 68 & 39\\
S190510g &  T+2.0 & 977 & 1166 (L) & 0.59 & 0.67 & 33 & & & 5 & 28\\
S190718y &  T+3.8& 368 & 7246 (B) & 0.17 & 0.22 & 45 & & & 27 & 18\\
S190728q &  T+12.7 & 144 & 543 (B) & 0.0063 & 0.0060 & 3 & & &  & 3\\
S190808ae$^{R}$  & T+3.4 & 36 & & & & 2 &  & & & 2\\
S190814bv/GW190814 & T+3.2 & 352 & 23 (L) & 0.83  & 0.90 & 94 & & 2 & 60 & 32\\
S190822c$^{R}$ &  T+2.0 & 37 \\
S190930t &  T+2.1 &  746 & 24220 (B) & 0.006 & 0.029 & 16 & & & 5 & 11\\ 
\hline
O3b\\
\hline
S191110af$^{R}$ &  T+2.9 & 797  & & & & 17 & & & 6 & 11\\
S191213g  & T+39.7 & 4 & 4480 (L) & 1.1~$\times$~10$^{-4}$ & N/A  \\
S191216ap  & T+6.1 & 113 & 253 (L) & 0.021 & 0.037 & 20 & & & 14 & 6\\
S200114f & T+1.8 & 206 & 403 (cWB) & 0.30 &   & 8 & 1 & & 6 & 1\\
S200115j$^\dagger$  & T+2.0 & 512 & 765 (L) & 0.047 & 0.097 & 82 & & 9 & 41 & 32\\
S200213t & T+5.7 & 9 & 2326 (L) & 2.1~$\times$~10$^{-4}$ & 1.1~$\times$~10$^{-4}$ & 5 & & & 5  \\
S200224ca &  T+6.1 & 670  & 72 (L) & 0.82 & 0.80  & 8 & & & 2 & 6\\
S200225q &  T+47.9 & 70 & 22 (L) & 0.51 & 0.51 & 1 & & & & 1\\
\hline
\end{tabular}

\footnotesize{$^{R}$ Trigger was subsequently retracted and no longer thought to be astrophysical.\\ $^\dagger$ The X-ray background for observations of S200115j was strongly elevated due to temperature issues, which caused a large number of spurious sources, many with a high rank. See text for more details.}
  
\label{table:triggers}
\end{center}
\end{table*}

\begin{figure*}
\begin{center}
  \includegraphics[clip, width=8.5cm]{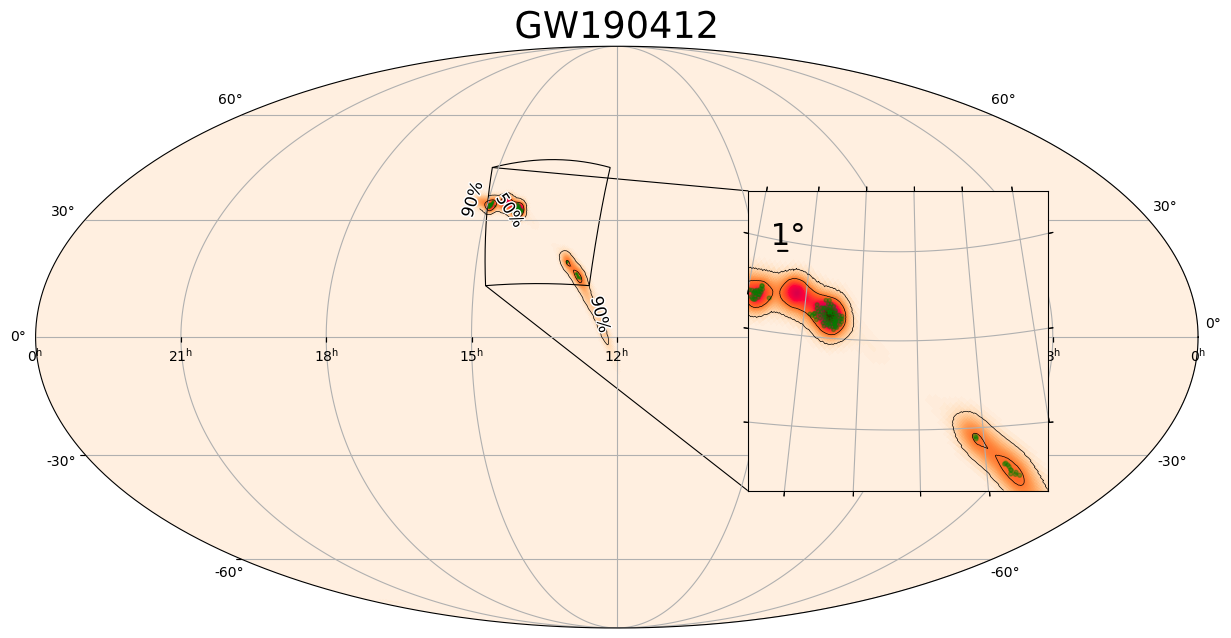}
  \includegraphics[clip, width=8.5cm]{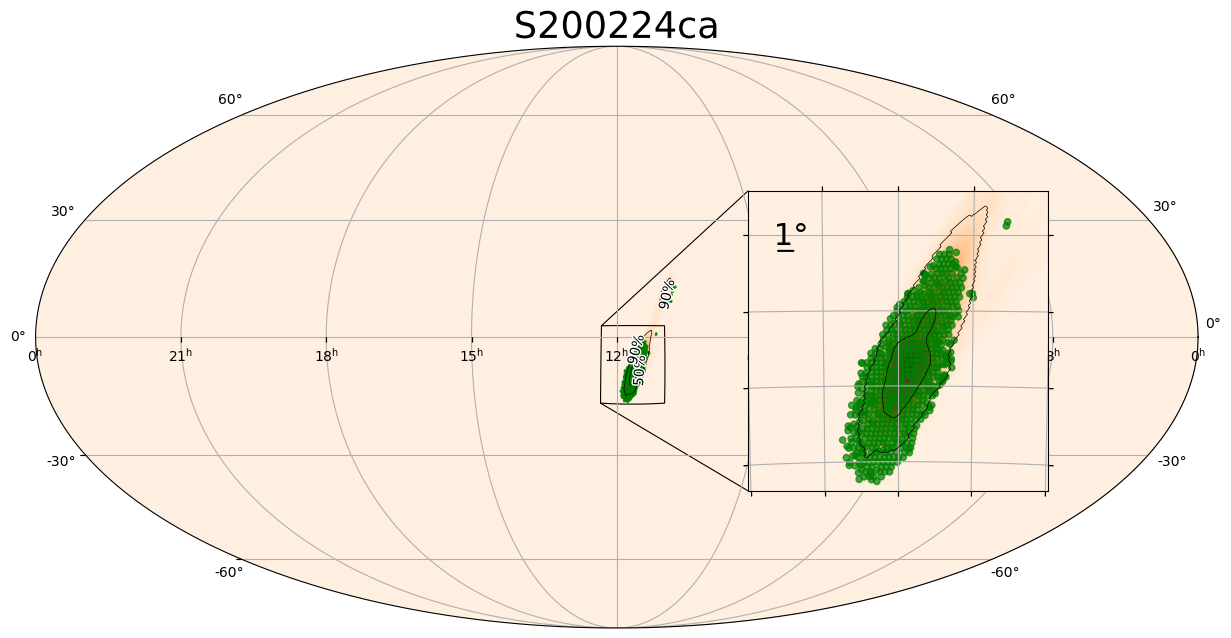}
  \includegraphics[clip, width=8.5cm]{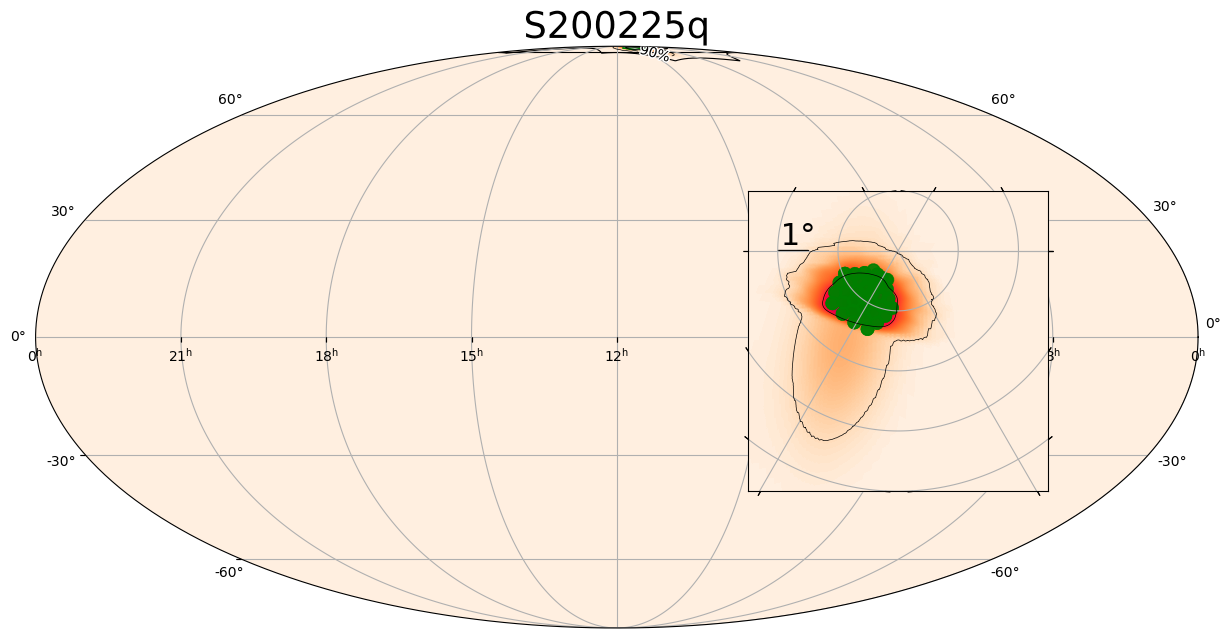}
\caption{Tiling observations for the BBH triggers. 90~per~cent and 50~per~cent GW error contours are shown by the solid lines; green spots mark the footprints (location and size) of the XRT fields observed. In the case of S200225q, the fields are all concentrated around the northern pole close to 90 deg. in declination, hence a polar view is shown in the inset. S191216ap is not shown, since follow-up of the IceCube error region was performed instead of standard tiling. }
\label{bbhtiles}
\end{center}
\end{figure*}

\begin{figure*}
\begin{center}
  \includegraphics[clip, width=8.5cm]{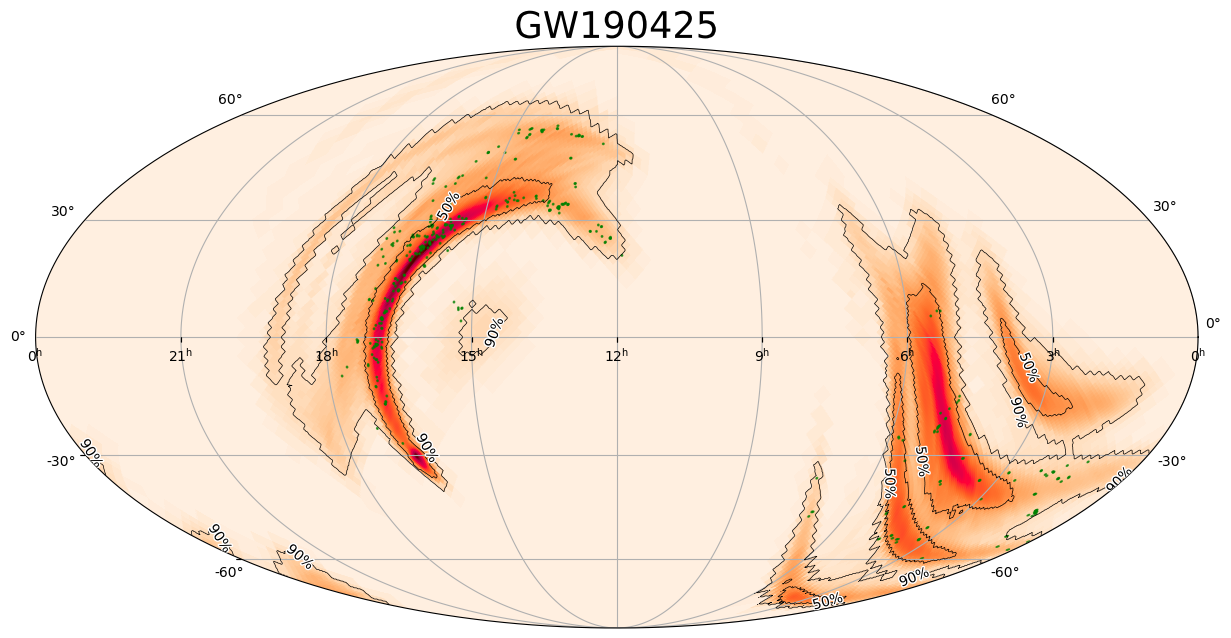}
  \includegraphics[clip, width=8.5cm]{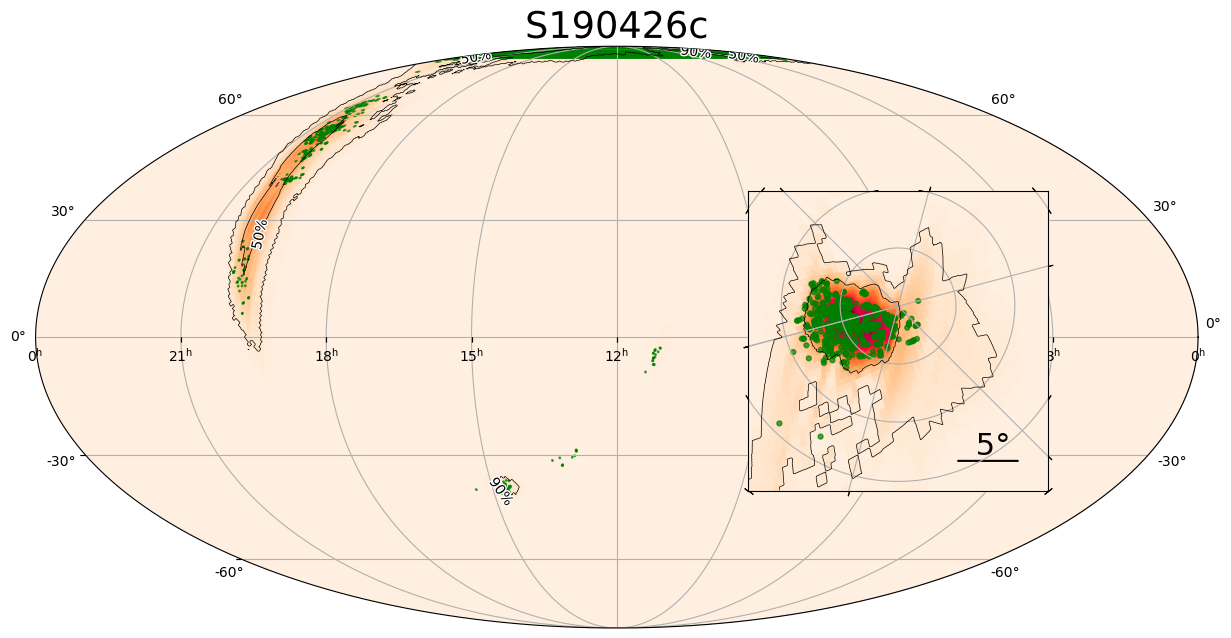}
  \includegraphics[clip, width=8.5cm]{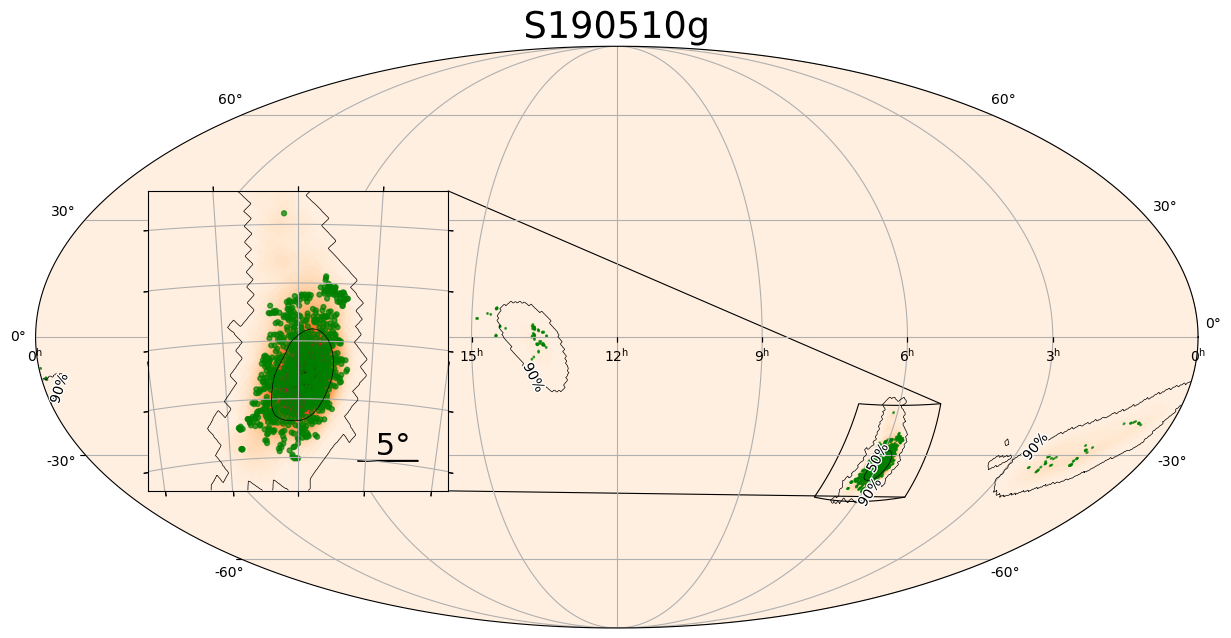}
  \includegraphics[clip, width=8.5cm]{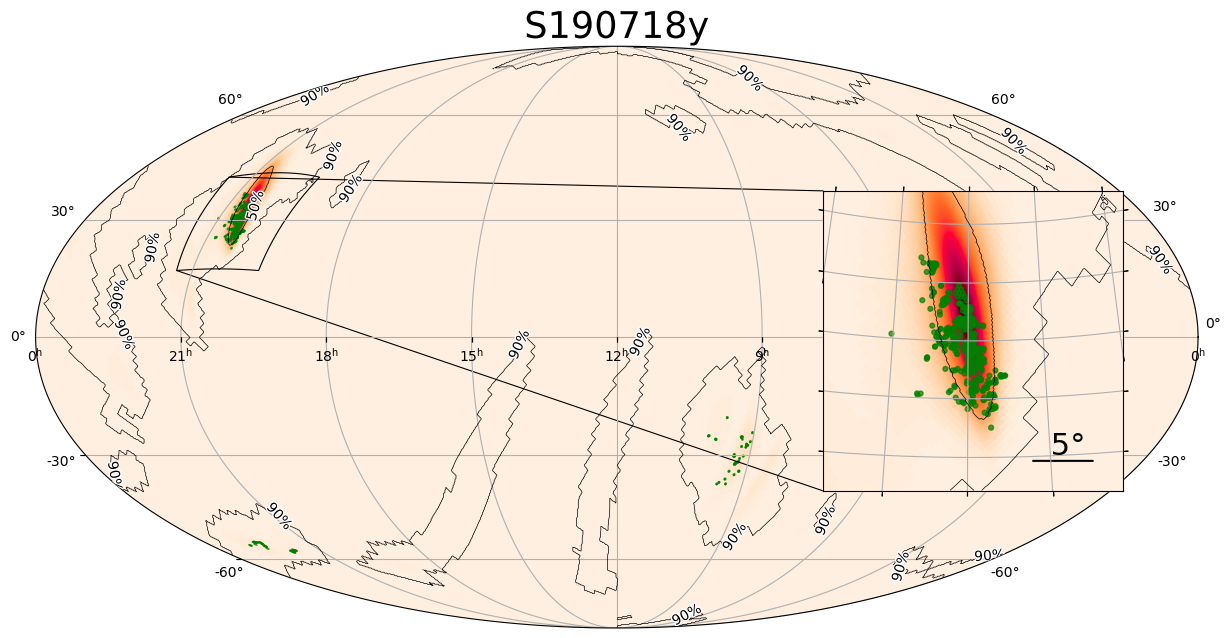}
\caption{Tiling observations for the BNS triggers. 90~per~cent and 50~per~cent GW error contours are shown by the solid lines; green spots mark the footprints (location and size) of the XRT fields observed. The inset for S190426c shows a view of the region around 90 deg. declination. Not shown are S191213g and S200213t, for which follow-up of individual externally-discovered sources was performed, rather than a tiling pattern.}
\label{bnstiles}
\end{center}
\end{figure*}

\begin{figure*}
\begin{center}
  \includegraphics[clip, width=8.5cm]{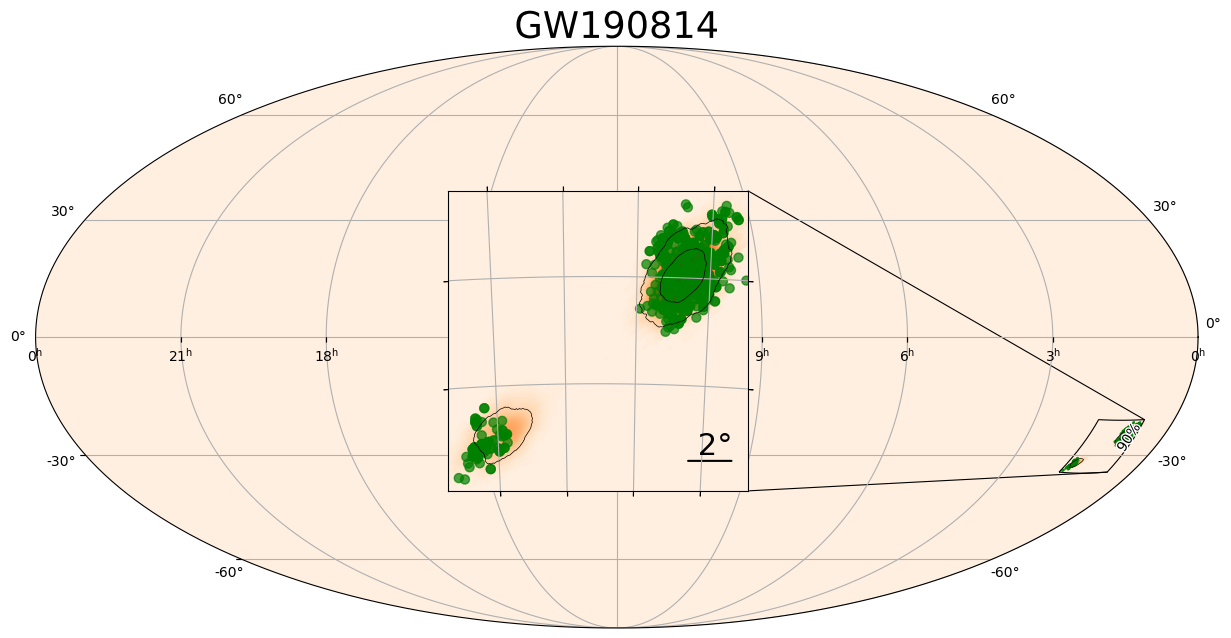}
  \includegraphics[clip, width=8.5cm]{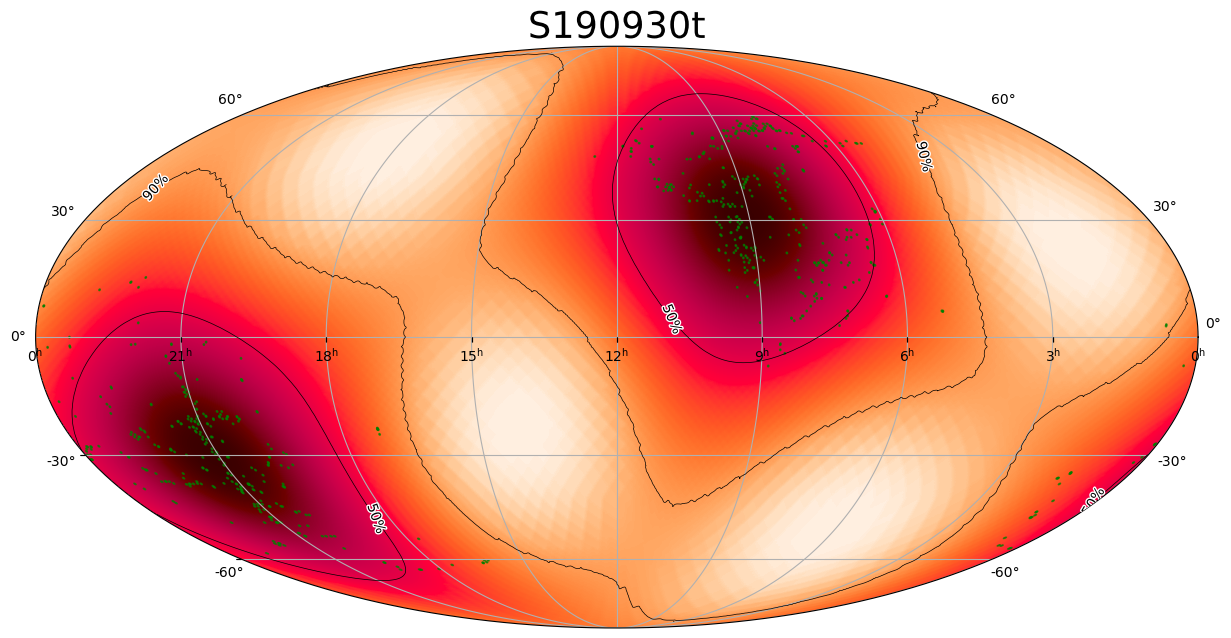}
  \includegraphics[clip, width=8.5cm]{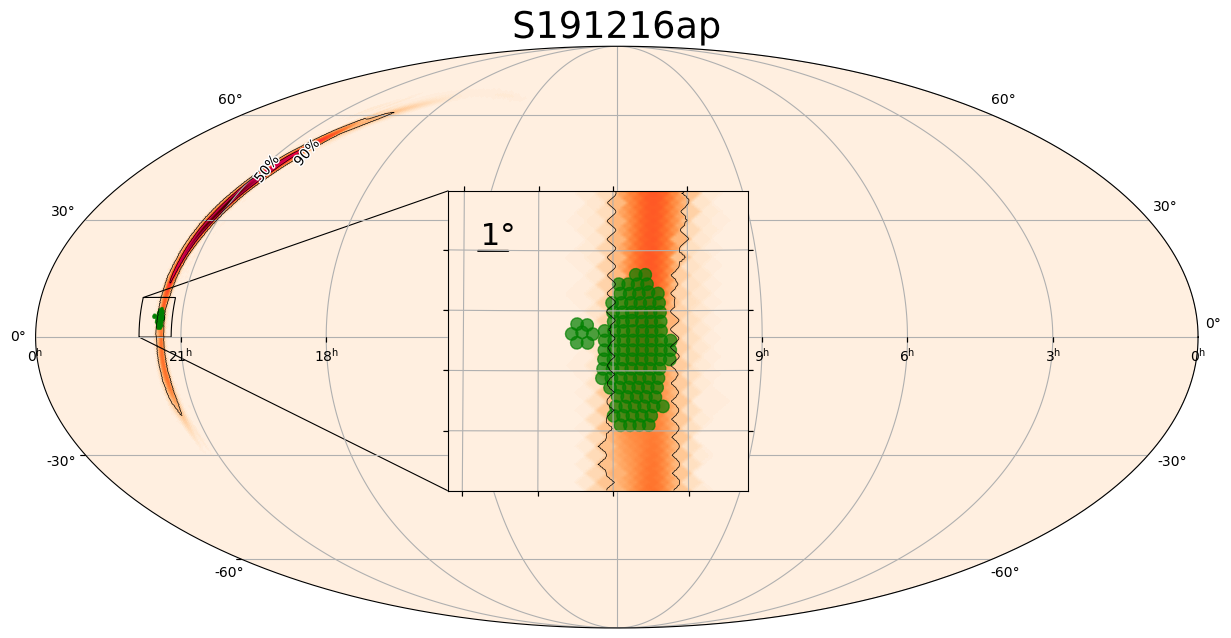}
  \includegraphics[clip, width=8.5cm]{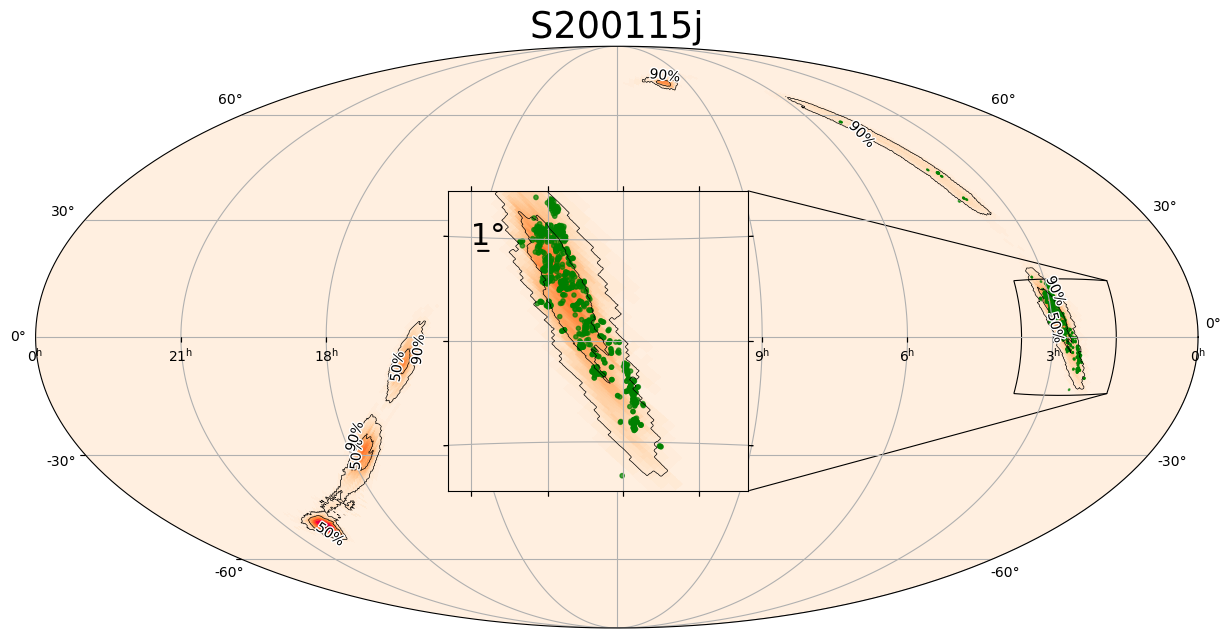}
  \includegraphics[clip, width=8.5cm]{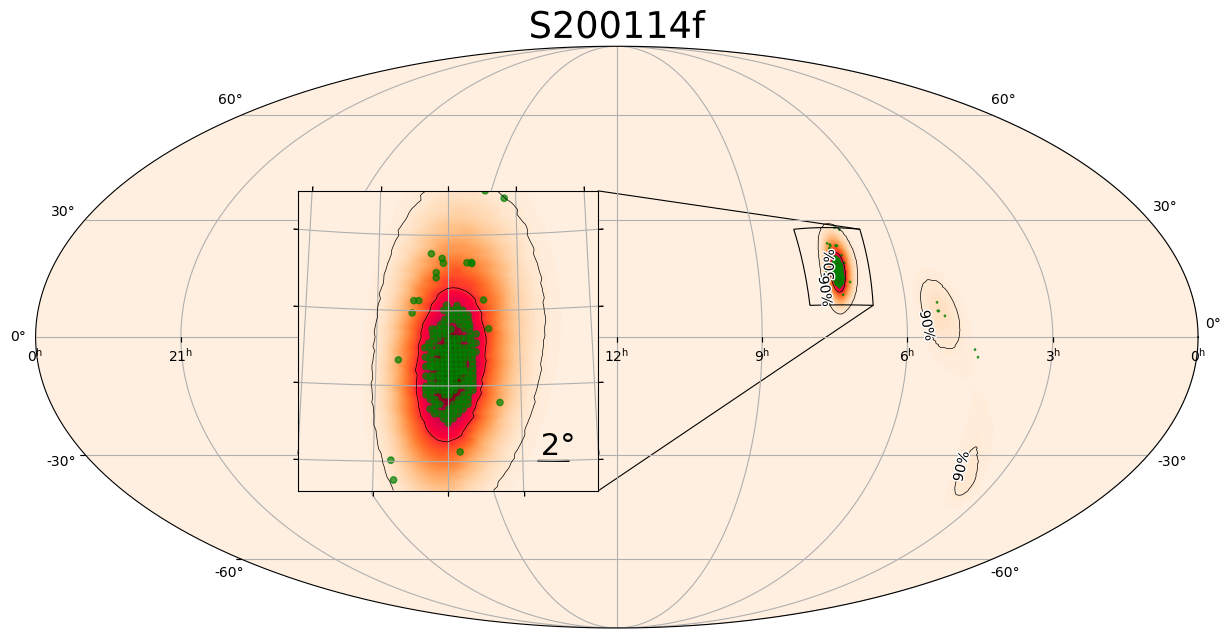}
\caption{Tiling observations for the NSBH (top row), Mass Gap (middle row) and unmodelled (bottom) triggers. 90~per~cent and 50~per~cent GW error contours are shown by the solid lines; green spots mark the footprints (location and size) of the XRT fields observed.}
\label{othertiles}
\end{center}
\end{figure*}

\section{Results}
\label{results}

Here we consider each of the O3 GW triggers followed up by {\em Swift},
reporting on the actual observations performed and the X-ray sources
detected in each case. For completeness, even the triggers later
retracted have been included. Full details of the fields
observed can be found under each specific trigger page at
https://www.swift.ac.uk/LVC/, including {\em Swift} target IDs, start
time of the observation and the exposure for each
pointing. Information about each confirmed X-ray source found is also
provided there. In addition, the webpages list the X-ray count rate upper
limits for UVOT-detected sources with a good Q0 or Q1 quality flag
(Oates et al. in prep); given the large number of these sources, we do
not list them here, and only refer explicitly to those sources which
were deemed worthy of further {\em Swift} follow-up. The Gravitational Wave Treasure Map tool\footnote{http://treasuremap.space/} \citep{tm}, designed to visualise and coordinate EM follow-up, also includes detailed information about the {\em Swift} pointings. 

\subsection{S190412m (GW 190412)}

S190412m, a BBH merger at 812~$\pm$~194 Mpc, did not satisfy our criteria for follow-up: the area enclosing 90~per~cent of the probability was $\sim$~151.7~deg$^2$, significantly larger than our chosen cut of 10~deg$^2$. However, the trigger was used to test a new implementation of the tiling software. 94 pointings of $\sim$~80~s each were performed, covering 9~deg$^2$, spanning 38--62.6~ks after the LVC trigger (Fig.~\ref{bbhtiles}). Only three X-ray sources were detected, of which two were Rank 3 (uncatalogued X-ray sources with nothing to distinguish them from typical faint background sources; no further follow-up was performed)\footnote{Short follow-up observations of three UVOT candidates were also performed more than a year later in 2020 June; these sources were not detected by XRT, to $\lo$~0.03 count~s$^{-1}$.} and one Rank 4, correponding to a known Seyfert galaxy. 

Following offline analysis, S190412m was confirmed as a highly significant GW detection and renamed GW~190412 \citep{190412paper}.

\subsection{S190425z (GW 190425)}

S190425z was identified as a BNS merger ($>$99~per~cent probability; distance of 156~$\pm$~41 Mpc), so was immediately marked for follow-up. The GW candidate was poorly localised since it was essentially a single interferometer trigger (below threshold for the Virgo detector, and the LIGO-Hanford observatory was offline at the time). 403 observations, covering 46.7 deg$^2$ on the sky, were performed between 16--59~ks after the trigger. This covered 6.5~per~cent of the LALInference skymap, after convolving with the 2MPZ galaxy catalogue (Fig.~\ref{bnstiles}). For this GW event, the initial 80~s tiles were observed, together with target of opportunity (ToO) observations, each of between 1 and 3~ks for four specific sources reported as candidate counterparts in the optical band: ZTF19aarykkb, ZTF19aarzaod \citep{24191}, AT2019ebq \citep[=PS19qp;][]{24210} and Swift J170219-122908 \citep[this last source being a possible transient found in an earlier UVOT observation;][]{24296}; these observations occurred 36.5--220.6~ks after the GW trigger. There was also a late time (2020 June) observation of a possible UVOT counterpart (Swift J065827.6-454319.8).
None of these ToO sources was detected by the XRT, to limits of between 2.5~$\times$~10$^{-3}$ and 0.02 count~s$^{-1}$. ZTF19aarykkb, ZTF19aarzaod and AT2019ebq were subsequently classified as Type II or Ib/IIb supernovae \citep{24204, 24205, 24233}, and therefore unrelated to the GW trigger.

Follow-up 500-s observations were not performed because of S190426c, another GW trigger the following day, taking precedence.

In total, nine X-ray sources were identified: two Rank 3, and seven Rank 4. Of the Rank 4 sources, four are classified as active galactic nuclei (AGN), one a galaxy in a cluster and two were previously listed in {\em ROSAT} and {\em XMM-Newton} catalogues, at approximately the same flux level as measured here. Of the sources marked as Rank 3, one is identified as a galaxy, and one as an AGN, both classifications based on optical data, with no previous X-ray observations reported.

S190425z is also now officially known as GW~190425 \citep{190425paper}.

\subsection{S190426c}

S190426c had P$_{\rm BNS}$~=~0.49 and P$_{\rm NSBH}$~=~0.12, strongly indicating the merger involved a NS; the distance estimate was 376~$\pm$~100 Mpc. Four months later, \cite{25549} updated the classification of this source to have an increased probability of 0.58 of being terrestrial in origin (c.f. 0.14 when the trigger was first announced). Tiling of 894 unique fields was carried out (both 80-s and 500-s exposures, leading to 1048 pointings in total; Fig.~\ref{bnstiles}), running between 8.6-548~ks after the trigger, detecting 107 X-ray sources, of which 68 were Rank 3, and 39 Rank 4 (7 AGN, 18 with no classification other than `X-ray source' -- mainly from {\em ROSAT}, 4 galaxies and 10 stars of different types). The observations covered 67~deg$^2$ on the sky, equating to 31~per~cent of the LALInference skymap after galaxy convolution.

As well as the standard tiling routine, pointed observations of ZTF19aassfws \citep{24331} were performed around 1675~ks (20 days) after the trigger, though the source was not detected by XRT (to a 3$\sigma$ upper limit of 1.7~$\times$~10$^{-3}$ count~s$^{-1}$); this source has since been retracted as a possible counterpart \citep{kasliwal20b}. In addition, XRT source 5 (= 1SXPS~J144850.8$-$400845) in the field was chosen for more follow-up, since it was highlighted as being a factor of $\sim$~5.3 brighter than an earlier, serendipitous observation of that field in 2011 \citep{24273}. However, the source is ranked as level 4, a known X-ray source consistent with being an AGN. This slight brightening is therefore most likely to be an AGN flare (see Section~\ref{disc} for further discussion of AGN activity). Swift~J201946.1+594818, a source detected in the initial UVOT tiling, was also re-observed as a potential counterpart, though subsequently decided not to be of interest \citep{24863}; it was not detected by XRT down to 3.9~$\times$~10$^{-3}$ count~s$^{-1}$.

\subsection{S190510g}

S190510g initially had a high (0.97) chance of being a BNS system, at a distance of 227~$\pm$~92~Mpc. However, the classification was updated the following day to P$_{\rm BNS}$~=~0.42 and P$_{\rm Terres.}$~=~0.58. {\em Swift} observed 977 fields (mainly $\sim$~80~s tiles) between 7.2--270~ks, covering 76.9~deg$^2$ on the sky (corresponding to 67~per~cent of the probability in the galaxy-convolved LALInference skymap; Fig.~\ref{bnstiles}). Follow-up was aborted once the classification of a BNS merger became less likely. 33 X-ray sources were detected, with all but five being previously known (the remaining five were Rank 3). Of the 28 Rank 4 sources, nine are classified as AGN, 11 as galaxies (or a cluster of galaxies), six are different types of stars, and the remaining two are unknown X-ray sources in {\em ROSAT} catalogues.

\subsection{S190718y}

This trigger had a 0.97 probability of being a terrestrial noise event; however, were it to be real, then the probability was that it was formed through a binary NS merger; following the initial decision tree, any GW events which were flagged as containing a NS would be followed up. Therefore, 368 pointings were performed (both 80-s, where some were repeated, and 500-s exposures) from 13--365~ks after the trigger, covering 30.9~deg$^2$ (22~per~cent of the probability in the BAYESTAR skymap after convolution; Fig.~\ref{bnstiles}). If this event were cosmological, its distance is estimated to be 226~$\pm$~164 Mpc. A total of 45 X-ray sources were found, with 27 Rank 3 and 18 Rank 4. Of these known sources, three are AGN, five galaxies, five are unidentified {\em ROSAT} X-ray sources, four are stars, and one has no associated classification.

\subsection{S190728q}

Because of the large positional uncertainty from the LVC, the BBH S190728q (at a distance of  785~$\pm$~212 Mpc) did not pass the {\em Swift} filtering criteria for follow-up. However, IceCube \citep{icecube} announced a neutrino candidate \citep{25192, 25197, 25210}; with this better (though still large: radius of 4.8~deg) possible localisation, follow-up observations were approved (exposure times of $\sim$~100~s per tile), and ran for 45--104~ks after the LVC trigger, covering 14.3~deg$^2$ on the sky \citep[only 0.6~per~cent of the galaxy-convolved BAYESTAR skymap, though $\sim$~20~per~cent of the IceCube localisation; ][]{25220}. Within these observations, three Rank 4 sources were identified, corresponding to an active galaxy, a star and an infrared (IR) source. 

A ToO observation of ZTF19abjethn/AT2019lvs \citep{25199} was also performed, finding a corresponding X-ray source at a level of 2.9$^{+1.4}_{-1.1}$~$\times$~10$^{-3}$ count~s$^{-1}$. This source was subsequently noted to be outside the updated skymap, and thus retracted as a potential counterpart \citep{25207}; \cite{25204} and \cite{25209} also classified it as a cataclysmic variable.

\subsection{S190808ae - Retracted}

S190808ae was initially announced as a CBC trigger with P$_{\rm BNS}$~=~0.42 and P$_{\rm Terres.}$~=0.57. A retraction was issued by the LVC around 6~hr later \citep{25301}, by which time a series of 80-s observations had already begun with {\em Swift}. In total, 36 pointings were performed between 12--19~ks, detecting two known galaxies.

\subsection{S190814bv (GW 190814)}

While S190814bv was initially classified as a Mass Gap trigger with a large area (370~deg$^2$ of the convolved skymap enclosing the 90~per~cent probability, P$_{\rm 0.9}$), an updated BAYESTAR skymap about 1.5~hr later decreased this error substantially such that $\sim$~18~deg$^2$ would enclose P$_{\rm 0.9}$. While this still did not satisfy the standard {\em Swift} follow-up criteria (Table~\ref{table:rules}), a judgement call was made to observe anyway. In addition, the classification was updated the following day to NSBH, together with a further refinement of the error region  with a LALInference skymap \citep[and distance of 267~$\pm$~51 Mpc;][]{25333}, further supporting our planned follow-up, given that the system likely contained a NS. Significant ground-based follow-up was performed of this trigger \citep[e.g., ][]{ackley20, thakur20, andreoni20}.

 {\em Swift} observed 352 fields from 11-471~ks (both initial 80-s tiles, many of which were repeated, and later 500-s pointings; in total, 529 observations were taken); these covered 20.3~deg$^2$, corresponding to 89~per~cent of the galaxy-convolved LALInference skymap (Fig.~\ref{othertiles}), or 78~per~cent of the earlier BAYESTAR map. In addition to the standard tiling algorithm, observations were planned to cover rising radio source and optical transient AT~2019osy (also known as ASKAP~005547$-$270433) which had been announced \citep{25487, 25488, dobie19}; this source was not detected in X-rays \citep[upper limit of 2.0~$\times$~10$^{-3}$ count~s$^{-1}$; also undetected by {\em Chandra}:][]{25822}, with only the nearby galaxy visible in the UVOT data \citep{25545}. In total, 94 X-ray sources were detected, with 32 Rank 4, 60 Rank 3 -- and two Rank 2 sources (that is, sources marked as possible afterglows), sources 2 and 99 in the field.

Upon further investigation, sources 2 and 99 were found to be the same source, not correctly aggregated due to a bug in the automated analysis software.  They were highlighted as potentially interesting due to being almost 9$\sigma$ (a flux ratio of 2.9) brighter than the catalogued count rate in the 1SXPS catalogue (1SXPS J005446.7-245528 = 2SXPS J005446.7-245530). Follow-up observations initially suggested a possible fading trend, though the latest data, collected 13.5~days after the trigger, showed the source to have rebrightened again. While not a previously catalogued X-ray source, the position is consistent with a known AGN which could easily explain the variability. This source is clearly detected in the UVOT data, with an AB magnitude of $u$~$\sim$~17.1--17.2 during the observations. 
 
A number of the Rank 3 and 4 sources in the field were also considered of possible interest, due to their potential variability, and were therefore re-observed by {\em Swift}. None of the follow-up observations of the Rank 3 sources (sources 14, 31, 51, 59, 74 and 88) identified significant fading, so they were dropped from further consideration.

Source 6 matches XMMSL2 J005323.1-244018  in the {\em XMM-Newton} slew catalogue, though is significantly fainter ($\sim$~0.1~count~s$^{-1}$ compared with the catalogued rate of $\sim$~0.8~count~s$^{-1}$ -- both in terms of XRT counts). There was a slight indication that the source was rising, so additional observations were taken. However, the count rates remained consistent within the error bars.

Source 7, a previously catalogued {\em ROSAT} source called 1RXS J005355.4-240439, was 2.6$\sigma$ (flux ratio of 5.2) brighter than the catalogued level, and also fading at 3.2$\sigma$. This fading is noted between a single point and all the later measurements which are consistent with the {\em ROSAT} measurement. This source is also catalogued as a possible AGN.

Source 43, also detected as a bright source by {\em ROSAT} (1RXS J005040.5-254115), faded at 2.0$\sigma$ between one higher measurement (consistent with the catalogued rate) and the subsequent fainter detections. This source is a known AGN.

Of the 32 known Rank 4 sources (including those discussed explicitly above), ten are classified as AGN, seven each as galaxies and {\em ROSAT} X-ray sources of unspecified type, two are stars, one is an ultra-luminous X-ray source, one a supernova remnant, with the final four catalogued as unknown types of X-ray sources detected by other missions. 

{\em Swift} observations of S190814bv will be described in more detail by Cenko et al. (in prep.). Following confirmation of its reality as a merger of a BH with some form of compact object (either the lightest black hole or the heaviest NS yet discovered), at 241$^{+41}_{-45}$~Mpc, this trigger is also known as GW~190814 \citep{190814paper}.

\subsection{S190822c - Retracted}

After an initial classification as a BNS, prompting {\em Swift} observations, S190822c was retracted as an astrophysical source about 45~min later \citep{25442}. However, before this updated information, a 37 point tiling observation (80~s per tile) had been uploaded to the spacecraft, and ran from 7.3--15~ks after the trigger. An AGN and a previously unknown X-ray source were found in these data.  

\subsection{S190930t}

S190930t was marked as 74~per~cent likely to be a NSBH merger at a distance of 108~$\pm$~37 Mpc; the LVC error region was very large, covering the majority of the sky, since it was a single interferometer detection. {\em Swift} observed 735 different fields between 7.6--120~ks, covering 83.1~deg$^2$ of the sky, and 2.9~per~cent of the convolved BAYESTAR skymap (Fig.~\ref{othertiles}). After the initial 80-s tiles, further observations were performed of a number of new sources identified in the early XRT (one source of interest) and UVOT (ten sources) data, as well as the externally-detected AT2019rpn \citep[also known as ZTF19acbpqlh; ][]{25899}, which was not detected in XRT observations, down to an upper limit of 5.3~$\times$~10$^{-3}$ count~s$^{-1}$; this source was later classified as a type II supernova, unrelated to S190930t \citep{kasliwal20b}. A final UVOT candidate was observed in 2020 June, but was undetected in XRT data ($<$0.025 count~s$^{-1}$).

From the XRT perspective, source 12 was initially considered as potentially interesting because it was detected at about twice the RASS upper limit (though with a substantial error bar) and faded slightly (at the 1.4$\sigma$ level) between the initial observation at $\sim$~34~ks after the GW trigger and later observations starting at $\sim$~240~ks \citep{25966}. Beyond this time, the count rate remained around 0.01~count~s$^{-1}$. Again, despite not having a previous X-ray detection, the position is coincident with a known AGN.

Of the ten UVOT sources flagged to be followed up, one was thought to correspond to source 28 in the XRT list; this only produced an  unconstraining upper limit in the initial 80~s snapshot, but was then detected in longer follow-up observations, with a count rate varying between 0.009 and 0.03~count~s$^{-1}$. This source, seen to be fading in the UVOT, was subsequently named  AT2019sbk \citep{25964}. With additional data, the localisation of the X-ray source was improved, and determined to be unrelated to AT2019sbk, but instead consistent with the centre of the galaxy 2MASX~J22471856-5814422 \citep{25984}. 

None of the other potentially interesting UVOT sources was detected by XRT, to a typical upper limit of $\sim$~4--5~$\times$~10$^{-3}$ count~s$^{-1}$. At the time of writing (around 300 days post-trigger), Swift J221951$-$484240 \citep[a candidate transient identified in the UVOT data; ][]{25901, 25939} is still being regularly observed (Oates et al. in prep); given the larger amount of data, the X-ray upper limit for this source is deeper: 2.6~$\times$~10$^{-4}$ count~s$^{-1}$.

From all the tiling observations of this trigger, five Rank 3 and 11 Rank 4 sources were detected, of which five are known AGN, three are stars, one is a galaxy and the remaining two are previously catalogued {\em ROSAT} X-ray sources, showing no sign of outburst.

\subsection{S191110af - Retracted}

The first O3b trigger {\em Swift} observed was retracted four days after the event \citep{26250}. Initially classified as an unmodelled trigger with a central frequency of $\sim$~1.8~kHz, this was flagged as a possible Galactic event, so 80-s {\em Swift} observations were planned. After convolving the error region with the Galactic plane given the trigger type, 798 fields were observed, from 10--203~ks after the trigger, finding six Rank 3 and 11 Rank 4 sources. Unsurprisingly given the concentration around the Galactic plane, five of these sources were marked as stars, and two as high-mass X-ray binary systems. There were also two catalogued, though unidentified, X-ray sources and two known to be IR emitters.

\subsection{S191213g}
\label{191213g}

Although S191213g was marked as likely to be a BNS merger (76~per~cent; distance of 200~$\pm$~80 Mpc), the large error region meant that tiling by {\em Swift} would only cover 0.017 of the area in 24~hours, well below the chosen limit of 0.1. While no tiling was therefore performed, observations of three ZTF sources \citep[ZTF19acykzsk, ZTF19acyldun, ZTF19acymixu;][]{26424, 26437} and a Pan-STARRS candidate \citep[PS19hgw/AT2019wxt;][]{26485} did take place over the following few days \citep{26471, 26501}. None of these sources was detected by the XRT, with limits of $\sim$~4.5--6.5~$\times$~10$^{-3}$ count~s$^{-1}$, and were all later classified as supernovae unrelated to the GW event \citep{kasliwal20b, 26591}

\subsection{S191216ap}

S191216ap was a Mass Gap trigger, with a low (though non-zero) probability of hosting a disrupted NS, at 375~$\pm$~70 Mpc. The error region was large, meaning {\em Swift} tiling would only have covered $\sim$~0.33 of the area in 24~hr, whereas the follow-up criteria require  P$_{\rm 24 hr}$~$\go$~0.5. However, IceCube announced a counterpart neutrino candidate \citep{26460,26463}, and {\em Swift} performed 100 tiles (of $\sim$~50-60~s each) to cover the convolution of the neutrino and GW error regions, spanning 22--42~ks after the trigger and covering 10.2~deg$^2$ on the sky \citep[5.8~per~cent of the BAYESTAR skymap after convolution with the galaxy catalogue, and 65~per~cent of the probability contained within the combined GW and neutrino localisations; Fig.~\ref{othertiles}; ][]{26475}. 
In addition to this, HAWC \citep[High-Altitude Water Cherenkov observatory; ][]{hawc} detected a sub-threshold event with a position similar to that of the IceCube one, though not covered by the initial {\em Swift} tiling \citep{26472}. Therefore, a further 7-point tiling pattern (500~s per tile; this fully covered their 68~per~cent containment region) was observed, as well as specific pointings towards the nine galaxies mentioned by \cite{26479} as being coincident with the LIGO/Virgo and HAWC positions \citep{26498}. In total, 20 XRT sources were found: 14 Rank 3 and 6 Rank 4. Of the six previously-known sources, one is an AGN, three are galaxies and the remaining two are associated with known radio and IR/UV sources.

\subsection{S200114f}

S200114f was a low-frequency ($\sim$~65~Hz) unmodelled trigger. Given that the error region was relatively small, 80-s {\em Swift} observations were planned, with 206 tiles spanning 6.6--99~ks after the trigger; the trigger S200115j on the following day then took precedence. This covered 21.9~deg$^2$ on the sky, corresponding to 30~per~cent of the probability in the cWB skymap (Fig.~\ref{othertiles}). The BAT FOV covered almost 98~per~cent of the LVC probability \citep{26748} for this trigger; no counterpart candidates were identified.

Eight X-ray sources were found, including one flagged as a strong candidate to be the EM counterpart to the GW trigger (initially a Rank 2 source, later promoted to Rank 1 as the error bars on the integrated flux improved with more data); the others consisted of six Rank 3 and a single Rank 4 source (a rotationally-variable star also detected by {\em ROSAT}).

The source of interest (`source 2') was so flagged because it showed early indications of fading \citep{26787, 26791}. However, this source is spatially coincident with a known AGN. Repeated follow-up observations of the source were taken by {\em Swift}, to investigate its evolution. After the initial brief fading from 0.1 to 0.02 count~s$^{-1}$ around 52--53~ks after the trigger, the source stayed consistently $\sim$~0.06 count~s$^{-1}$; this was still the case when the source became too close to the Sun for {\em Swift} to observe, more than four months after the trigger. The corresponding UVOT source also showed no signs of variability, with $u$~$\sim$~16.9.

We note that the new Gamma-ray Urgent Archiver for Novel Opportunities \citep[GUANO;][]{aaron20} system for {\em Swift}-BAT was activated by the S200114f event, leading to good limits of $<$~8.1~$\times$~10$^{-8}$ erg~cm$^{-2}$~s$^{-1}$ (8$\sigma$ confidence level; 14--195~keV) being placed on a BAT prompt gamma-ray detection within $\pm$15~s of the GW trigger.

{\em Swift} follow-up of S200114f, including source 2, will be analysed in more detail by Evans et al. (in prep.).

\subsection{S200115j}

S200115j was classified as a Mass Gap event, with a high probability of containing a disrupted NS, at a distance of 340~$\pm$~79 Mpc. {\em Swift} observations covered 512 unique fields spanning 7.1--501~ks after the trigger: both the initial phase of 80-s observations (some of them repeated), and the longer 500-s exposures (for most of the fields) were performed, leading to 719 pointings in total. The localisation skymap changed considerably between the initial BAYESTAR and later LALInference maps, with the error region shifting and decreasing in size; the {\em Swift} observations were planned and initiated when only the BAYESTAR maps were available. In total, 36.2~deg$^2$ of the sky were covered, corresponding to 9.7~per~cent of the galaxy-convolved updated LALInference skymap. Fig.~\ref{othertiles} shows the tiles plotted over the LALInference map.

During the {\em Swift} observations, the XRT experienced an extended interval of higher than normal operating temperature. This led to increased instrumental background, and the issuing of automatic GCN notices for spuriously high-ranked sources \citep[Ranks 1 and 2;][]{26777}. As always, each source was vetted by a human, and any obviously spurious sources not promoted to the public page.

In total, XRT detected 82 sources we believe are likely to be real. Of these, nine are Rank 2 (sources of interest), 41 Rank 3, and 32 Rank 4. Looking into the apparently interesting sources in more detail \citep[see also ][]{26808, 26855}, it was found that sources 130, 488, 717 and 748 all correspond to (likely) AGN, while source 136 matches a 2MASS galaxy and 745 an emission line galaxy (Mrk 1036); additional observations over the following days and weeks showed nothing to distinguish them from ordinary AGN activity in these sources. There were corresponding UVOT $u$-band detections of each of these except source 488, but no significant evidence for variability in any of them.

Source 487 was noted as being above the RASS limit and fading, but this description is based on a single detection during the interval of high background (all other observations provided upper limits only), so is likely spurious; there was no detection by UVOT to a 3$\sigma$ limit of $u$~$>$~20--21. Source 707 was above the RASS detection limit, and faded between two detections (all other observations, out to 175 days after the trigger, were upper limits, suggesting possible further fading). Given that the source is relatively faint, even an extra photon or two from the high background could be skewing these results. Additionally, the source was originally only flagged as `reasonable' \cite[see ][ for a definition of the detection flags]{1sxps}, meaning that there is a $\sim$~7~per~cent probability of the source being spurious. There was no UVOT source detected at this location ($u$~$>$~21--22), although these observations did not start until 11 days after the GW trigger. In a similar vein, source 746 showed fading from above to below the RASS limit between two observations (although the second data point only contained three source counts; the background level is low enough that, using Bayesian statistics, this is still a strongly significant detection of $>$99.999~per~cent), with additional upper limits in between, and later limits (out to $\sim$~180 days post-trigger) implying the source had faded further. No counterpart was detected by UVOT to $u$~$>$~20.

Of the 32 Rank 4 known sources, 18 are AGN, 4 are galaxies, 3 are X-ray sources of an unknown type, 3 correspond to stars (including one RS~CVn type) and the remaining four are simply catalogued as IR, UV or `blue' sources.

\subsection{S200213t}
\label{200213t}

S200213t was a BNS trigger (distance of 200~$\pm$~80 Mpc), but with too large an area to satisfy {\em Swift} follow-up criteria. However, following the announcement of a neutrino candidate from IceCube \citep{27043}, a seven-point tiling plan was uploaded \citep[$\sim$~1.3 ks per tile; ][]{27121}. Additionally, ToO observations of ZTF20aanakcd \cite[also known as AT2020cmr; ][]{27068} and ZTF20aamvmzj \citep[also known as AT2020cja; ][]{27051}, possible optical counterparts to the GW event, were performed \citep{27153,27400}. In total nine separate fields were observed, running from 20--22.5~ks after the trigger (for the seven point tiling; observations of ZTF20aamvmzj/AT2020cja continued until a month after the trigger), and covering 1.2~deg$^2$ on the sky (0.01~per~cent of the galaxy-convolved LALInference map). While neither of the ToO sources was detected by the XRT ($<$3.1~$\times$~10$^{-3}$ and 6.7~$\times$~10$^{-4}$ count~s$^{-1}$ for ZTF20aanakcd and ZTF20aamvmzj/AT2020cja, respectively), five Rank 3 field sources were identified. ZTF20aanakcd was later found to be a type IIn supernova and unrelated to S200213t \citep{27075}.

\subsection{S200224ca}

S200224ca was announced as a BBH at a distance of 1574~$\pm$~322 Mpc. It was well localised, such that 90~per~cent of the convolved probability covered only 72~deg$^2$ of the sky (Fig.~\ref{bbhtiles}), meaning that P$_{\rm 24 hr}$~=~0.66 for {\em Swift}, within our monitoring criteria. {\em Swift} observed 670 separate fields, from 21--196~ks after the trigger, covering 64.5~deg$^2$ on the sky, corresponding to 79~per~cent of the probability in the convolved LALInference skymap. Most of the observations were the initial 80-s tiles; as the second phase of 500-s tiles was begun, a new GW trigger, S200225q, was announced, and took precedence. Six UVOT sources were also followed-up with further ToO observations, but all were undetected in X-rays (Oates et al. in prep).

Within the tiling observations, only eight X-ray sources were identified, two Rank 3 and six Rank 4 (all of which are consistent with catalogued AGN).

In 2020 May, the unknown X-ray sources 5 and 9 were further observed. Source 5 showed no real evidence for fading, whereas source 9 had faded between the observations performed 1.6 and 90--120 days after the GW trigger (the later observations providing upper limits; there were only four source counts in the initial detection, however, and the source is only flagged as `reasonable'). 

This trigger will be individually discussed in a future publication (Klingler et al. in prep.).

\subsection{S200225q}

S200225q, at a distance of 994~$\pm$~187 Mpc, was strongly (95~per~cent) associated with a BBH merger, although the FAR of 1/3.5~yr was higher than the cut-off chosen for standard {\em Swift} follow-up. However, given the very good updated localisation (50~per~cent area of only 3~deg$^2$, leading to P$_{\rm 24 hr}$~=~0.87 of the convolved map) released 38~hr later \citep{27229}, a decision was made to follow-up. The initial plan of performing a 37~point tiling was interrupted by a trigger on GRB~200227A \citep{27234}. Restarting the tiling later, a total of 70 observations of $\sim$~80~s each were performed, from 172--224~ks after the trigger, and covering 3.8~deg$^2$ on the sky. This covered 51~per~cent of the galaxy-convolved LALInference skymap (Fig.~\ref{bbhtiles}). Because the follow-up was interrupted and delayed, the second phase of 500~s tiles was not performed. Only a single X-ray source was detected \citep{27526}, corresponding to a previously-catalogued {\em ROSAT} object.

\section{Discussion}
\label{disc}

During the third LIGO/Virgo observing run, {\em Swift} followed up 18 of the GW triggers announced, three of which were subsequently retracted, performing almost 6500 separate pointings. Of the 15 non-retracted triggers, four were classified as likely BBHs, six as BNS mergers, two NSBHs, two Mass Gap events and one an unmodelled (Burst) trigger. In total, four O3 triggers (GW 190412, 190425, 190521 and 190814) have been officially confirmed as being real GW events. While {\em Swift} detected many X-ray sources during these observations, none stands out as a likely EM counterpart to a GW event -- that is, a new bright (or significantly brightened) source, close to a known galaxy (see description of source rankings in Section~\ref{swift}).

\subsection{Sources}
\label{agn}

Much of the cosmic X-ray background can now be resolved as emission from discrete AGN \citep[e.g.][]{shanks91, barcons07, caccianiga08, mateos08, corral14, bat105}, and such active galaxies are inherently variable at X-ray (and other) wavelengths, over timescales from as short as minutes up to many years \citep[e.g., ][]{mchardy85, barr86, mushotzky93, bbf96, giommi19}. It is therefore unsurprising that wide-field observations such as those performed in the follow-up of large GW error regions reveal variable X-ray sources, under the assumption that many of these unknown sources are AGN. Considering the 18 LVC (15 likely real, three retracted) triggers followed-up by {\em Swift} during O3, 198 catalogued (Rank 4) sources were found. Of these, a third (66) are known AGN. In addition, of the 11 unique sources which were flagged as potentially interesting (i.e., Rank 1 or 2), eight correspond to AGN (the remaining three were uncatalogued). A total of 243 previously uncatalogued Rank 3 sources were detected, and it is very likely that many of these will be AGN. Of the remaining Rank 4 sources, the next largest population corresponds to galaxies (some in clusters), with 34 of the sources being classified as such.

\cite{190521gpaper} report a possible EM counterpart for the (probable) BBH merger S190521g (not followed-up by {\em Swift} because of the large area), which is consistent with the merger occurring in, and interacting with, the accretion disc of an AGN. This suggests that perhaps more attention should be paid to following up AGN in the GW error regions in future runs; knowledge of whether emission from a given active galaxy typically remains close to constant, before showing an {\em unusual} flaring event close in time to a GW trigger \citep[as was the case for J124942.3+344929/ZTF19abanrhr in][]{190521gpaper} would be useful, though may be difficult to achieve.

\subsection{Flux distribution}

\begin{figure}
\begin{center}
  \includegraphics[clip, width=8.5cm]{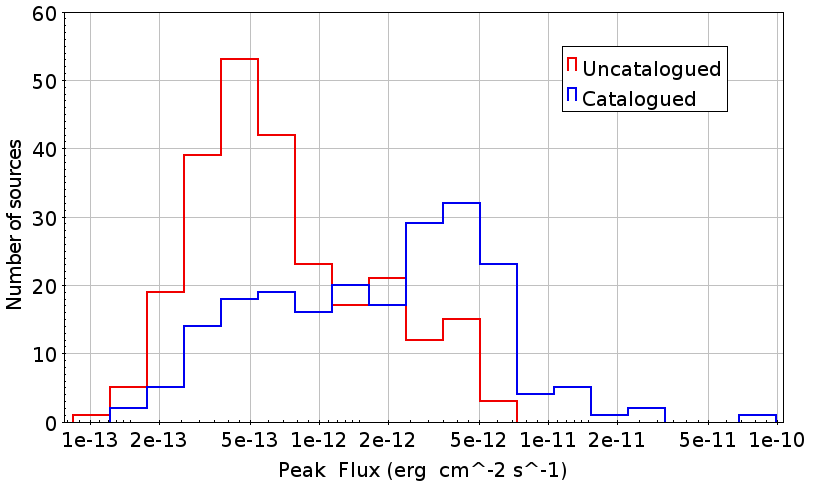}
 \caption{Histogram of the peak X-ray fluxes (0.3-10 keV) for the uncatalogued and catalogued X-ray sources across all GW follow-up during O3.}
\label{sources}
\end{center}
\end{figure}

Fig.~\ref{sources} shows histograms of the peak fluxes of the uncatalogued and catalogued sources detected by the XRT during the O3 observing run. These 0.3--10~keV absorbed fluxes are estimated from the measured peak count rate by assuming a power-law spectrum with photon index $\Gamma$~=~1.7 and an absorbing column of N$_{\rm H}$~=~3~$\times$~10$^{20}$~cm$^{-2}$ (a conversion factor of $\sim$~4.3~$\times$~10$^{-11}$ erg~cm$^{-2}$~count$^{-1}$). Note that the fluxes plotted are the {\em peak} values from the observations.

Unsurprisingly, the uncatalogued sources are skewed towards lower fluxes than the catalogued ones, with the majority of the sources detected having an observed 0.3--10~keV flux of around 5~$\times$~10$^{-13}$ erg~cm$^{-2}$~s$^{-1}$; the previously catalogued source number peaks about an order of magnitude brighter than this. The median values are more similar, at $\sim$~9.9~$\times$~10$^{-13}$ and 1.7~$\times$~10$^{-12}$ for uncatalogued and catalogued sources, respectively.
The uncatalogued sources range in flux from  1.0~$\times$~10$^{-13}$ to 6.6~$\times$~10$^{-12}$ erg~cm$^{-2}$~s$^{-1}$, while the known sources cover 1.6~$\times$~10$^{-13}$ to 9.6~$\times$~10$^{-11}$ erg~cm$^{-2}$~s$^{-1}$, factors of $\sim$~60 and 600 between brightest and faintest, respectively. Only one of the catalogued sources is brighter than 5~$\times$~10$^{-11}$erg~cm$^{-2}$~s$^{-1}$, though: source 1 in the field of S190510g, which is known to be a quasar (QSO~B0548$-$322). Excluding this source, the ratio of brightest to faintest catalogued sources is $\sim$~200.

In comparison, the 2SXPS catalogue has a median 0.3--10~keV flux of 4.7~$\times$~10$^{-14}$ erg~cm$^{-2}$~s$^{-1}$, more than a factor of ten lower; however, the mean exposure time for an observation in 2SXPS is $\sim$~2~ks, much longer than the $\sim$~80--500~s exposures obtained during the GW tiling \citep{2sxps}.

\subsection{Detection limits}
\label{det}

\begin{figure}
\begin{center}
  \includegraphics[clip, width=10cm]{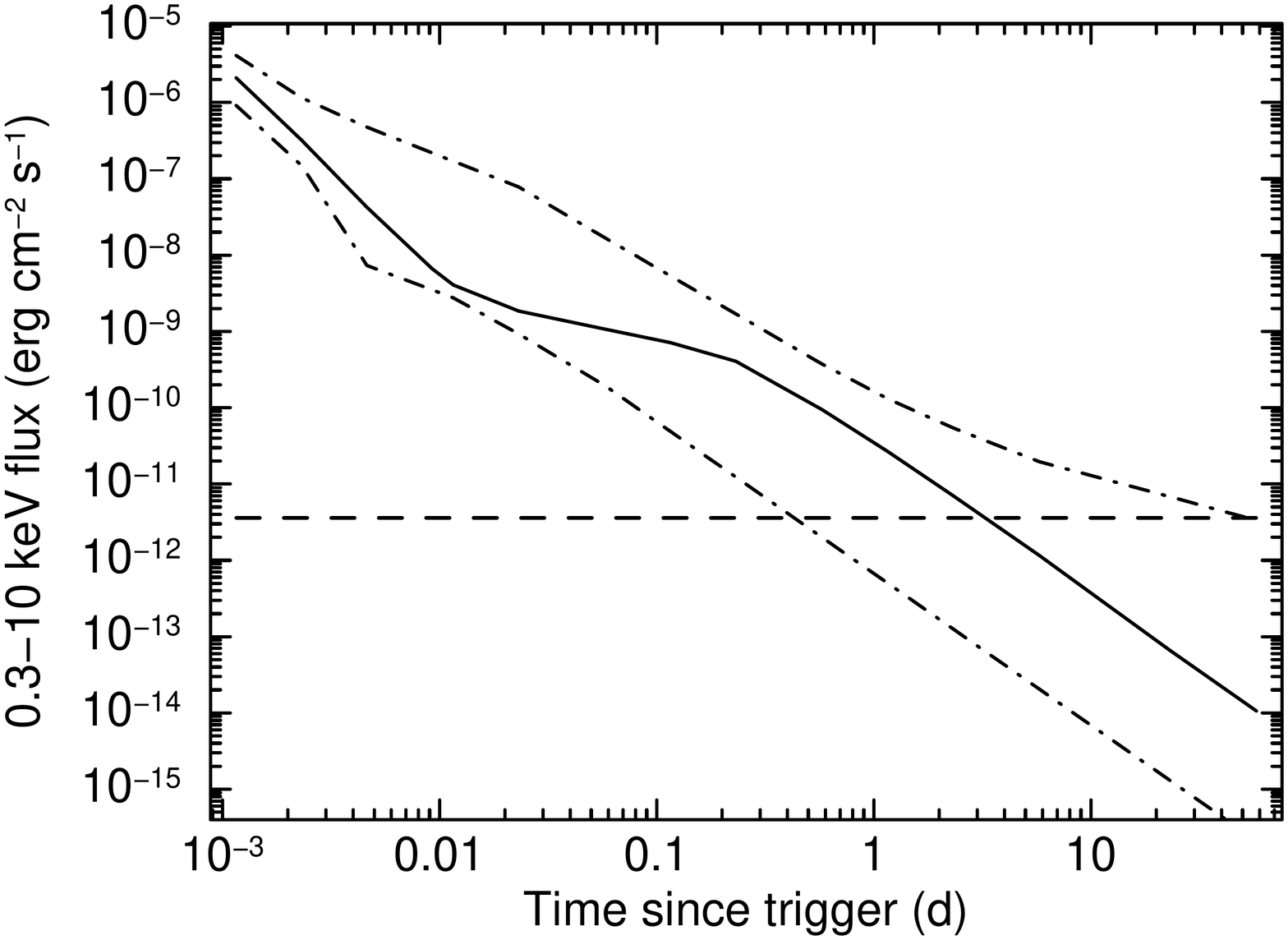}
\caption{X-ray afterglow light-curve for on-axis short GRBs, scaled to 140 Mpc. The solid black line shows the median curve, while the dot-dashed lines mark the 25th and 75th percentiles. The grey horizontal dashed line indicates the average flux upper limit discussed in $\S$~\ref{det}.}
\label{detlim}
\end{center}
\end{figure}

No strong candidates for X-ray counterparts to any of the O3 GW triggers were identified. While different triggers were observed for different amounts of time (though the majority of fields were observed for $\sim$~80~s), it may still be instructive to provide typical upper limits on source detections. For each (non-retracted) GW event where the standard large-scale tiling was performed, we estimate the mean 3$\sigma$ upper limit on the X-ray count rate by averaging the XRT non-detections for the UVOT Q0/Q1 sources for that trigger; since the UVOT sources are scattered throughout the area covered, this should provide a good estimate of the limiting brightness for X-ray sources in each GW error region. These values are given in Table~\ref{table:limits}; the same conversion factor as above from count to flux units was used. The average 0.3--10 keV flux upper limit across all the fields is 3.60~$\times$~10$^{-12}$ erg~cm$^{-2}$~s$^{-1}$. The O3 GW triggers (BBH, NSBH and BNS combined) span a large range of estimated distances, from 108 to 1574~Mpc (S190930t and S200224ca, respectively), with a mean value of 474~Mpc. For BNS events -- for which a short GRB-like counterpart is most likely -- the GW network sensitivity was $\sim$~140 Mpc\footnote{See
\url{https://www.gw-openscience.org/detector\_status/day/20200227/}, where the last part of the URL can be replaced with any date during
O3, in the format YYYYMMDD.}, hence we scaled GRB light curves to this distance in Fig~\ref{detlim}.
The corresponding average luminosity upper limit is therefore $\sim$~10$^{44}$~erg~s$^{-1}$ (for a distance of 474~Mpc), or $\sim$~10$^{43}$~erg~s$^{-1}$ (at 140~Mpc).

Figure~\ref{detlim} shows the median flux light-curve for short GRBs based on the flux-limited sample of \cite{paolo14}, shifted to a distance of 140~Mpc, and plotted against days since trigger. For GRBs without a measured redshift, the average value of $z$~=~0.85 was assumed. This plot shows that, for a typical on-axis short GRB, we would expect the X-ray afterglow to be above 3.60~$\times$~10$^{-12}$ erg~cm$^{-2}$~s$^{-1}$ for around 3$^{+50}_{-2.5}$~d after the trigger, and therefore readily detectable by XRT observations.

\begin{table}

\caption{X-ray upper limits (0.3--10~keV) for GW triggers where large-scale tiling patterns were performed. The last column gives the approximate fraction of the (updated, if relevant) galaxy-convolved skymap observed by {\em Swift}.}

\begin{center}
\begin{tabular}{lccc}
\hline
LVC trigger &   3$\sigma$ UL &Obs. flux UL & Fraction of\\
& (count s$^{-1}$) & (erg cm$^{-2}$ s$^{-1}$) & skymap covered\\
\hline
S190412m & 0.087 & 3.7~$\times$~10$^{-12}$ & 0.17\\
S190425z & 0.15 & 6.5~$\times$~10$^{-12}$ & 0.065\\
S190426c & 0.046 & 2.0~$\times$~10$^{-12}$ & 0.31\\
S190510g & 0.091 & 3.9~$\times$~10$^{-12}$ & 0.67\\
S190718y & 0.025 & 1.1~$\times$~10$^{-12}$ & 0.22\\
S190728q & 0.070 & 3.0~$\times$~10$^{-12}$ & 0.006\\
S190814bv & 0.023 & 9.9~$\times$~10$^{-13}$ & 0.90\\
S190930t & 0.10 & 4.3~$\times$~10$^{-12}$ & 0.03\\
S191216ap & 0.19 & 8.2~$\times$~10$^{-12}$ & 0.037\\
S200114f  &  0.092 & 4.0~$\times$~10$^{-12}$ & 0.30\\
S200115j &  0.021 & 9.0~$\times$~10$^{-13}$ & 0.097\\
S200224ca  & 0.093 & 4.0~$\times$~10$^{-12}$ & 0.80\\
S200225q & 0.10 & 4.3~$\times$~10$^{-12}$ & 0.51\\
\hline
\end{tabular}

\label{table:limits}
\end{center}
\end{table}

\section{Future prospects}

A significant challenge for the {\em Swift} follow-up thus far has been the fact that more than half of the sources detected during O3 were Rank 3; that is, we cannot tell whether they are new sources, or just too faint to have been previously detected. As mentioned earlier, the ongoing SGWGS observations (75~per~cent complete at the end of 2020 July) provide snapshot observations with which to compare later X-ray detections (something which is done automatically by the analysis software), which helps to mitigate the problem to some extent. 
More importantly, however, {\em eROSITA} \citep[extended ROentgen Survey with an Imaging Telescope Array; ][]{predehl17, merloni20}, launched in 2019, will perform a new all-sky survey over the next few years, covering an energy range comparable to {\em Swift}-XRT. With expected soft (0.5--2~keV) and hard (2--10~keV) all-sky sensitivities of $\sim$~4.4~$\times$~10$^{-14}$ and $\sim$~7.1~$\times$~10$^{-13}$ erg~cm$^{-2}$~s$^{-1}$ (for point sources), respectively, over the first six months, deepening to $\sim$~1.1~$\times$~10$^{-14}$ and $\sim$~1.6~$\times$~10$^{-13}$ erg~cm$^{-2}$~s$^{-1}$ after the full four years \citep{predehl17}, {\em eROSITA} will be able to detect, or place deeper upper limits on, many previously uncatalogued sources. 
More precisely, if we consider the Rank 3 sources identified by XRT throughout the O3 run, and use a range of power-law indices ($\Gamma$ = 0.5--2.0, with a typical absorbing column of N$_{\rm H}$~=~3~$\times$~10$^{20}$~cm$^{-2}$) to estimate the corresponding fluxes, the {\em eROSITA} four year sensitivity should allow the survey to detect all of these sources over one or both energy bands. The sources detected by XRT would thus be classed as Rank 4 (catalogued) instead, if they were at a consistent or fainter flux than the earlier {\em eROSITA} detection; if brighter by at least 3$\sigma$, the source would be promoted to Rank 2 and marked for additional follow-up. This would decrease the number of candidate counterparts by about 95~per~cent, significantly improving our ability to highlight the potentially interesting sources\footnote{We note that the proprietary period for the German {\em eROSITA} data will be two years.}. 
 
We note that analysis by \cite{bz20} suggests that, while the {\em eROSITA} survey will significantly increase the number of X-ray detected normal (inactive) galaxies, this will still only be a few per~cent of the total population (considering galaxies at a distance of 50--200~Mpc).

Besides {\em eROSITA}, {\em Einstein Probe} \citep{ep}, aimed for launch by the end of 2022, has a large FOV of 3600 deg$^2$ ($\sim$~1~sr), and will observe the whole sky over 0.5--5~keV at high cadence, detecting X-ray transients with which {\em Swift} detections in GW follow-up observations can be compared. The ECLAIRs coded-mask detector onboard {\em SVOM} \citep[Space-based multi-band astronomical Variable Objects Monitor;][]{svom}, due to be launched in 2021, has a 2~sr FOV with an energy bandpass down to 4~keV, and is expected to detect $\sim$~70 GRBs per year, adding to the chance that a short GRB coincident with a GW trigger will be detected.

As the sensitivity of the GW network improves, with more triggers at ever increasing distances, the incompleteness of galaxy catalogues will become more of a complication. Unless additional sensitive interferometers are included in the network, the positional errors will still remain large. However, a fourth LIGO interferometer in India is planned for the future, and KAGRA sensitivity should improve over the next few years. Selecting which triggers to follow, and optimising galaxy catalogues (which is currently being performed by a number of different groups in different ways), will be key to maximising the probability of detecting an EM counterpart.

\section{Data availability}

The data underlying this article are available in the {\em Swift} archives at https://www.swift.ac.uk/swift\_live/, \\https://heasarc.gsfc.nasa.gov/cgi-bin/W3Browse/swift.pl and \\https://www.ssdc.asi.it/mmia/index.php?mission=swiftmastr, with the relevant target IDs provided at https://www.swift.ac.uk/GW/.

\section*{ACKNOWLEDGEMENTS}
\label{ack}

KLP, PAE, APB, AAB, NPMK, JPO, CP and MJP acknowledge funding from the UK Space Agency. NJK acknowledges support from NASA Grant 80NSSC19K0408.
AD acknowledges financial contribution from the agreement ASI-INAF n.2017-14-H.0, while EA, MGB, SC, GC, AD, AM and GT acknowledge funding from the Italian Space Agency, contract ASI/INAF n. I/004/11/4. This work is also partially supported by a grant from the Italian Ministry of Foreign Affairs and International Cooperation Nr. MAE0065741 (AD), and by the Ministry of Education, Culture, Sports,
Science, and Technology (MEXT) KAKENHI Grant Numbers 17H06357 and 17H06362
(TS). DBM acknowledges research grant 19054 from VILLUM FONDEN.

This work made use of data supplied by the UK Swift Science Data Centre at the University of Leicester. We thank Dr Gavin Lamb for useful discussions.

\appendix
\section{O3 triggers}

Table~\ref{table:alltriggers} lists relevant information for all the LVC triggers from the third observing run, including those which were subsequently retracted, noting which were followed-up by {\em Swift}. 

 \onecolumn
\begin{landscape}
 
\begin{longtable}{lllllll}  
  \caption{LVC O3 triggers. If one classification is $\go$90~per~cent likely, that is the only one given in the third column; otherwise percentages are provided in parentheses. Terres. means Terrestrial (i.e. noise). P$_{\rm 24 hr}$ refers to the probability of the LIGO error region observable in 24 hours by {\em Swift}. The final column lists the GCN circulars and papers published by the LVC or {\em Swift} teams.  }
  \label{table:alltriggers}\\
\hline
LVC trigger & Trigger time & Trigger type & {\em Swift} & Reason & Notes & LVC/{\em Swift}  \\
ID & & /classification & follow-up? & if no &  & GCN circs\\
\hline
\hline
\endfirsthead

\multicolumn{6}{c}%
{{\bfseries \tablename\ \thetable{} -- continued from previous page}} \\
\hline
LVC trigger & Trigger time & Trigger type & {\em Swift} & Reason & Notes &LVC/{\em Swift}\\
ID & UTC & /most prob. class. & follow-up? & if no & & GCN circs.\\
\hline
\hline
\endhead

\hline \multicolumn{6}{|r|}{{Continued on next page}} \\ \hline
\endfoot

\hline \hline
\endlastfoot
O3a\\
\hline
S190405ar &	2019-04-05 16:01:30 & CBC/Terres. & N &  RETRACTED & & \cite{24109}\\
\hline
S190408an &	2019-04-08 18:18:02 & CBC/BBH & N & Area too large & & \cite{24069, 24075}\\
\hline
S190412m &	2019-04-12 05:30:44 & CBC/BBH & Y &  & Did not satisfy {\em Swift}  & \cite{24098, 24114}\\
& & & & & follow-up criteria, but used & \cite{190412paper}  \\
& & & & & to test new tiling software.\\
\hline
S190421ar &	2019-04-21 21:38:56 & CBC/BBH  & N & Area too large & & \cite{24141, 24158}; \\
& & & & & & \cite{24375}\\
\hline
S190425z &	2019-04-25 08:18:05 & CBC/BNS & Y & & & \cite{24168, 24184};\\
  & & & & & & \cite{24228, 24296}; \\
& & & & & & \cite{24305}\\
  & & & & & & \cite{24767, 190425paper}\\
\hline
S190426c &	2019-04-26 15:21:55 & CBC/BNS & Y & & & \cite{24237, 24255};\\
& & & & & & \cite{24273, 24277, 24279}\\
& & & & & & \cite{24353}; \\
& & & & & &  \cite{24411, 24863};\\
 & & & & & &  \cite{25549}\\
\hline
S190503bf &	2019-05-03 18:54:04 & CBC/BBH & N & Area too large & & \cite{24377, 24391}\\
\hline
S190510g &	2019-05-10 02:59:39 & CBC/Terres. (58);  & Y & & & \cite{24442, 24448, 24454} \\
& &BNS (42) & & & & \cite{24462, 24489, 24541};\\
& & & & & &  \cite{24862} \\

\hline
S190512at &	2019-05-12 18:07:14 & CBC/BBH & N & Area too large & &\cite{ 24503, 24518}; \\
& & & & & & \cite{24584}\\
\hline
S190513bm &	2019-05-13 20:54:28 & CBC/BBH & N & Area too large & & \cite{24522, 24543}\\
\hline
S190517h &	2019-05-17 05:51:01 & CBC/BBH & N & Area too large & & \cite{24570, 24582}\\
\hline
S190518bb &	2019-05-18 19:19:19 & CBC/BNS & N &  RETRACTED & & \cite{24591}\\
\hline
S190519bj &	2019-05-19 15:35:44 & CBC/BBH & N & Area too large & & \cite{24598, 24610}\\
\hline
S190521g &	2019-05-21 03:02:29 & CBC/BBH & N & Area too large & & \cite{24621, 24640, 24646}\\
& & & & & & \cite{190521paper}\\
\hline
S190521r &	2019-05-21 07:43:59 & CBC/BBH & N & Area too large & & \cite{24632, 24645}\\
\hline
S190524q &	2019-05-24 04:52:06 & CBC/Terres. (71); & N &  RETRACTED & & \cite{24656}\\
& & BNS (29) \\
\hline
S190602aq &	2019-06-02 17:59:27 & CBC/BBH & N & Area too large & & \cite{24717, 24725}\\
\hline
S190630ag &	2019-06-30 18:52:05 & CBC/BBH & N & Area too large & & \cite{24922, 24938};\\
& & & & & & \cite{25094}\\
\hline
S190701ah &	2019-07-01 20:33:06 & CBC/BBH & N & Area too large & & \cite{24950, 24966};\\
& & & & & & \cite{24987}\\
\hline
S190706ai &	2019-07-06 22:26:41 & CBC/BBH & N & Area too large & & \cite{24998, 25025};\\
& & & & & & \cite{25049}\\
\hline
S190707q &	2019-07-07 09:33:26 & CBC/BBH & N & Area too large & & \cite{25012, 25044}; \\
& & & & & & \cite{25048}\\
\hline
S190718y &	2019-07-18 14:35:12 & CBC/BNS & Y & & & \cite{25087, 25096}; \\
& & & & & & \cite{25107, 25151}\\
\hline
S190720a &	2019-07-20 00:08:36 & CBC/BBH & N & Area too large & & \cite{25115, 25127};\\
& & & & & & \cite{25138}\\
\hline
S190727h &	2019-07-27 06:03:33 & CBC/BBH & N & Area too large & & \cite{25164, 25174};\\
& & & & & & \cite{25249}\\
\hline
S190728q &	2019-07-28 06:45:10 & CBC/BBH & Y & &Large area, but follow-up of &  \cite{25187, 25200}; \\
& & & & & IceCube localisation & \cite{25208, 25220}\\
\hline
S190808ae &	2019-08-08 22:21:21 & CBC/Terres. (57); & Y & RETRACTED & & \cite{25296}; \cite{25301}\\ 
& &  BNS (43)\\
\hline
S190814bv &     2019-08-14 21:10:39 & CBC/Mass Gap  & Y &  & Small error  & \cite{25324, 25333, 25341}; \\
& &updated to NSBH & & & region & \cite{25400, 25525};\\
& & & & & & \cite{25545, 190814paper}\\
\hline
S190816i &	2019-08-16 13:04:31 & CBC/NSBH (83);  & N &  RETRACTED& & \cite{25367}\\
& & Terres. (17)\\
\hline
S190822c &	2019-08-22 01:29:59 & CBC/BNS & N & RETRACTED & & \cite{25442}\\
\hline
S190828j &	2019-08-28 06:34:05 & CBC/BBH & N & Area too large & & \cite{25497, 25504, 25532}; \\
& & & & & & \cite{25861}\\
\hline
S190828l &	2019-08-28 06:55:09 & CBC/BBH & N & Area too large & & \cite{25503, 25533};\\
& & & & & & \cite{25782}\\
\hline
S190829u &	2019-08-29 21:05:56 & CBC/Mass Gap  & N & RETRACTED & & \cite{25554}\\
\hline
S190901ap &	2019-09-01 23:31:01 & CBC/BNS & N & Very poor localisation & & \cite{25606}; \cite{25614}; \\
& & & & & & \cite{25617}\\
\hline
S190910d &	2019-09-10 01:26:19 & CBC/NSBH & N & Poor localisation  & & \cite{25695, 25704};\\
& & & & & & \cite{25723}\\
\hline
S190910h &	2019-09-10 08:29:58 & CBC/BNS (61);  & N & Very poor localisation  & & \cite{25707, 25718};\\
& &Terres. (39) & & & & \cite{25778}\\
\hline
S190915ak &	2019-09-15 23:57:02 & CBC/BBH & N & Area too large & & \cite{25753, 25765}; \\
& & & & & & \cite{25773}\\
\hline
S190923y &	2019-09-23 12:55:59 & CBC/NSBH (68);  & N & Poor localisation,   & & \cite{25814, 25846}\\
 & &Terres. (32) & & high FAR, high P(Terres.)\\
\hline
S190924h &	2019-09-24 02:18:46 & CBC/Mass Gap  & N & Behind the Sun & & \cite{25829, 25843};\\
& & & & & & \cite{25909}\\
\hline
S190928c &	2019-09-28 02:11:45 & Burst & N & RETRACTED & No prompt notices sent & \cite{25883}\\
\hline
S190930s &	2019-09-30 13:35:41 & CBC/Mass Gap  & N & Area too large & & \cite{25871, 25889}; \\
& & & & & & \cite{25968}\\
\hline
S190930t &	2019-09-30 14:34:07 & CBC/NSBH (74);  & Y & & & \cite{25876, 25888}; \\
& & Terres. (26)& & & & \cite{25901, 25939};\\
& & & & & & \cite{25964};\\
& & & & & & \cite{25966, 25984}\\
\hline
O3b\\	
\hline
S191105e &	2019-11-05 14:35:21 & CBC/BBH & N & FAR too high & & \cite{26182, 26192};\\
& & & & & & \cite{26245}\\
\hline
S191109d &	2019-11-09 01:07:17 & CBC/BBH & N & Area too large & & \cite{26202, 26211}\\
\hline
S191110x &	2019-11-10 18:08:42 & CBC/Mass Gap & N & RETRACTED & &\cite{26218}\\
\hline
S191110af &	2019-11-10 23:06:44 & Burst (Galactic) & N & RETRACTED & & \cite{26222, 26234}; \\
& & & & & & \cite{26238, 26250}\\
\hline
S191117j &	2019-11-17 06:08:22 & CBC/NSBH & N &  RETRACTED & & \cite{26254}\\
\hline
S191120aj &	2019-11-20 16:23:34 & CBC/NSBH (61);  & N &  RETRACTED & & \cite{26263}\\
& & Terres. (39)\\
\hline
S191120at &	2019-11-20 20:08:37 & CBC/Mass Gap (83);  & N & RETRACTED & & \cite{26265}\\
& & Terres. (17)\\
\hline
S191124be &	2019-11-24 09:59:18 & CBC/Mass Gap & N &  RETRACTED & & \cite{26288}\\
\hline
S191129u &	2019-11-29 13:40:29 & CBC/BBH & N & Area too large & & \cite{26303, 26313}; \\
& & & & & & \cite{26383}\\
\hline
S191204r &	2019-12-04 17:15:26 & CBC/BBH & N & Area too large & & \cite{26334, 26348}\\
\hline
S191205ah &	2019-12-05 21:52:08 & CBC/NSBH & N & P$_{\rm 24 hr}$ too low & & \cite{26350, 26365}\\
\hline
S191212q &	2019-12-12 08:27:28 & CBC/Terres. (51); & N &  RETRACTED & & \cite{26395}\\
& &  NSBH (49)\\
\hline
S191213g &	2019-12-12 04:34:08 & CBC/BNS (76);  & Y & -- & Only low P(24 hr) could be  & \cite{26402, 26410}; \\
& &Terres. (23) & & & covered but followed-up of & \cite{26417, 26471};\\
& & & & & externally-detected sources & \cite{26501}\\
\hline
S191213ai &	2019-12-13 15:59:05 & CBC/NSBH (85); & N & RETRACTED & & \cite{26413}\\
& & Terres. (15) \\
\hline
S191215w &	2019-12-15 22:30:52 & CBC/BBH & N & Area too large & &\cite{26441, 26453};\\
& & & & & & \cite{26518}\\
\hline
S191216ap &	2019-12-16 21:33:38 & CBC/Mass Gap & Y & & P$_{\rm 24 hr}$ low, but &  \cite{26454, 26466}; \\
& & & & &followed up IceCube and & \cite{26475, 26498, 26505};\\
& & & & & HAWC regions & \cite{26570}\\
\hline
S191220af &	2019-12-20 12:24:14 & CBC/BNS & N &  RETRACTED & & \cite{26513}\\
\hline
S191222n &	2019-12-22 03:35:37 & CBC/BBH & N & P$_{\rm 24 hr}$ too low & &\cite{26543, 26557};\\
& & & & & & \cite{26572}\\
\hline
S191225aq &	2019-12-25 21:57:15 & CBC/Terres. (61);   & N & RETRACTED & & \cite{26585}\\
& &  Mass Gap (39)\\
\hline
S200105ae &	2020-01-05 16:24:26 & CBC/Terres.  & N & FAR too high & Sub-threshold but likely real & \cite{26640, 26642, 26649};\\

& & & & & & \cite{26657, 26675, 26688}\\
\hline
S200106au &	2020-01-06 18:34:29 & CBC/Terres. & N &  RETRACTED & & \cite{26641}\\
\hline
S200106av &	2020-01-06 18:34:23 & CBC/Terres. & N &  RETRACTED & & \cite{26641}\\
\hline
S200108v &	2020-01-08 10:00:38 & CBC/BBH  & N &  RETRACTED & & \cite{26665}\\
\hline
S200112r &	2020-01-12 15:58:38 & CBC/BBH & N & P$_{\rm 24 hr}$ too low & & \cite{26715, 26723}\\
\hline
S200114f &	2020-01-14 02:08:18 & Burst & Y & & & \cite{26734, 26748}; \\
& & & & & & \cite{26787, 26791}\\
\hline
S200115j &	2020-01-15 04:23:09 & CBC/Mass Gap & Y & & & \cite{26759, 26777}; \\
& & & & & & \cite{26779}; \\
& & & & & & \cite{26798, 26807};\\
& & & & & & \cite{26808, 26855}\\
\hline
S200116ah &	2020-01-16 11:56:42 & CBC/NSBH & N &  RETRACTED & & \cite{26785}\\
\hline
S200128d &	2020-01-28 02:20:11 & CBC/BBH & N & FAR too high & & \cite{26906, 26921}\\
\hline
S200129m &	2020-01-29 06:54:58 & CBC/BBH  & N &  P$_{\rm 24 hr}$too low & & \cite{26926, 26940}\\
\hline
S200208q &	2020-02-08 13:01:17 & CBC/BBH & N & P$_{\rm 24 hr}$ too low & & \cite{27014, 27023}; \\
& & & & & & \cite{27036}\\
\hline
S200213t &	2020-02-13 04:10:40 & CBC/BNS (63);  & Y & &  P$_{\rm 24 hr}$ low, but & \cite{27042, 27058}; \\
& &Terres. (37) & & & followed up IceCube and & \cite{27092, 27096, 27153, 27400};\\
& & & & & optical transients & \\
\hline
S200219ac &	2020-02-19 09:44:15 & CBC/BBH & N & FAR too high & & \cite{27130, 27147}; \\
& & & & & & \cite{27214}\\
\hline
S200224ca &	2020-02-24 22:22:34 & CBC/BBH & Y & & & \cite{27184, 27216}; \\
& & & & & & \cite{27262, 27288}; \\
& & & & & & \cite{27524}\\
\hline
S200225q &	2020-02-25 06:04:21 & CBC/BBH & Y &  & FAR high, but& \cite{27193, 27217}; \\
& & & &  & well localisaed& \cite{27229, 27526}\\
\hline
S200302c &	2020-03-02 01:58:11 & CBC/BBH (89); & N & P$_{\rm 24 hr}$ too low & & \cite{27278, 27289}; \\
& & Terres. (11) & & and FAR too high  & & \cite{27292}\\

\hline
S200303ba &	2020-03-03 12:15:48 & CBC/BBH (86);  & N &  RETRACTED & & \cite{27306}\\
& & Terres. (14)\\
\hline
S200308e &	2020-03-08 01:19:27 & CBC/NSBH (83); & N &  RETRACTED & & \cite{27347}\\
& &  Terres. (17)\\
\hline
S200311bg &	2020-03-11 11:58:53 & CBC/BBH & N & Behind the Sun & & \cite{27358, 27369}; \\
& & & & & & \cite{27382}\\	
\hline
S200316bj & 2020-03-16 21:57:56 & CBC/Mass Gap & N&  Initial P$_{\rm 24 hr}$ too low & & \cite{27388, 27401}\\
& & & &  Updated skymap improved & & \cite{27419} \\
& & & & error region, but not \\
& & & & until 5 days after trigger\\
\end{longtable}
\end{landscape}
\twocolumn

\bsp	
\label{lastpage}
\end{document}